\RequirePackage{cmap}

\documentclass{aa}
\makeatletter
\renewcommand*\aa@pageof{, page \thepage{} of \pageref*{LastPage}}
\makeatother

\usepackage[utf8]{inputenc}

\usepackage[T1]{fontenc}

\usepackage{amsmath}
\usepackage{amssymb}
\usepackage{bm}
\usepackage{textcomp}
\usepackage{upgreek}

\usepackage[varg]{txfonts}
\usepackage[scaled=0.76]{beramono}

\usepackage{fontawesome}

\usepackage[inline]{enumitem}

\usepackage{nicefrac}

\usepackage{csquotes}

\usepackage{microtype}

\usepackage{siunitx}
\DeclareSIUnit \magnitude {mag}
\DeclareSIUnit \parsec {pc}
\DeclareSIUnit \year {yr}
\DeclareSIUnit \pixel {pix} 
\usepackage{xcolor}

\PassOptionsToPackage{obeyspaces}{url}

\usepackage[
    unicode=true,
    colorlinks,
    urlcolor=blue,
    linkcolor=gray,
    citecolor=gray,
    filecolor=gray,
    pdfauthor={Timothy D. Gebhard, Markus J. Bonse, Sascha P. Quanz, Bernhard Schölkopf},
    pdftitle={},
    plainpages=false,
    pdfpagelabels,
]{hyperref}

\usepackage[capitalize]{cleveref}
\crefname{algocf}{Algorithm}{Algorithms}
\Crefname{algocf}{Algorithm}{Algorithms}
\crefname{section}{Sect.}{Sects.}
\Crefname{section}{Section}{Sections}

\usepackage[all]{hypcap}

\usepackage{natbib}
\bibpunct{(}{)}{;}{a}{}{,}

\usepackage{graphicx}

\usepackage{silence}
\usepackage{subcaption}

\usepackage{booktabs}

\usepackage[para]{threeparttable}

\usepackage{multirow}

\usepackage{adjustbox}

\usepackage{tabularx}

\usepackage{makecell}

\usepackage[algoruled, noalgohanging]{algorithm2e} 
\usepackage{standalone}

\usepackage{tikz}
\usetikzlibrary{
    shapes,
    calc,
    decorations.text,
    arrows,
    positioning,
    chains,
    quotes,
    shapes.geometric,
} 
\usepackage[]{datetime2}

\usepackage{lipsum}
\usepackage{kantlipsum}

\usepackage{verbatim}

\usepackage{xspace}

\usepackage{tcolorbox}
\newtcolorbox{extranote}{colback=blue!5!white, colframe=blue!20!white}

\usepackage{placeins}
\usepackage{afterpage} 
\newcommand{\eg}{e.g.,~}
\newcommand{\ie}{i.e.,~}

\newcommand{\indep}{\,\protect\rotatebox[origin=c]{90}{$\vDash$}\,}

\newcommand*\diff{\mathop{}\!\mathrm{d}}

\DeclareMathOperator*{\argmin}{arg\,min}

\usepackage{xstring}
\newcommand{\progid}[1]{\href{https://archive.eso.org/wdb/wdb/eso/sched_rep_arc/query?progid=#1}{\ttfamily#1}}
\newcommand{\simbad}[1]{\href{https://simbad.u-strasbg.fr/simbad/sim-basic?Ident=#1}{\StrSubstitute[0]{#1}{+}{~}}}

\makeatletter
\renewcommand{\object}[1]{%
   \global\let\if@myobjects\iftrue%
   \addcontentsline{obj}{obj}{#1}%
}
\makeatother

\newcommand{\logfpf}{\ensuremath{ \text{--\hspace{0.2pt}log}_\text{10}(\text{FPF})}\xspace }

\newcommand{\LinkToCode}{\small\faFileCodeO}

\newcommand{\Ypos}{\ensuremath{Y}\xspace}
\newcommand{\Yts}{\ensuremath{\mathcal{Y}}\xspace}
\newcommand{\Xpos}{\ensuremath{X}\xspace}
\newcommand{\Xts}{\ensuremath{\bm{\mathcal{X}}}\xspace}

\newcommand{\noop}[1]{}

\let\oldbibliography\thebibliography
\renewcommand{\thebibliography}[1]{%
  \oldbibliography{#1}%
  \setlength{\itemsep}{1.95pt}%
  \setlength{\baselineskip}{7pt}
  \setlength{\lineskiplimit}{-\maxdimen}
}

\begin{document}

    \title{
        Half-sibling regression meets exoplanet imaging:\\%
        PSF modeling and subtraction using a flexible,\\ domain knowledge-driven, causal framework
    }
    \titlerunning{Half-sibling regression meets exoplanet imaging}

    \author{
        Timothy D. Gebhard\inst{1,2,3}\fnmsep%
        \thanks{Correspondence: \href{mailto:tgebhard@tue.mpg.de}{tgebhard@tue.mpg.de}.} \and %
        Markus J. Bonse\inst{3} \and %
        Sascha P. Quanz\inst{3} \and %
        Bernhard Schölkopf\inst{1,4} %
    }
    \authorrunning{Gebhard et al.}

    \institute{
        Max Planck Institute for Intelligent Systems, Max-Planck-Ring 4, 72076 Tübingen, Germany \and %
        Max Planck ETH Center for Learning Systems, Max-Planck-Ring 4, 72076 Tübingen, Germany \and %
        ETH Zurich, Institute for Particle Physics \& Astrophysics, Wolfgang-Pauli-Str. 27, 8092 Zurich, Switzerland \and %
        Department of Computer Science, ETH Zurich, 8092 Zurich, Switzerland
    }

    \date{\textbf{Version:}~\DTMnow}

    \abstract{
        High-contrast imaging of exoplanets hinges on powerful post-processing methods to denoise the data and separate the signal of a companion from its host star, which is typically orders of magnitude brighter.
    }{
        Existing post-processing algorithms do not use all prior domain knowledge that is available about the problem. 
        We propose a new method that builds on our understanding of the systematic noise and the causal structure of the data-generating process.
    }{
        Our algorithm is based on a modified version of \emph{half-sibling regression} (HSR), a flexible denoising framework that combines ideas from the fields of machine learning and causality.
        We adapt the method to address the specific requirements of high-contrast exoplanet imaging data obtained in pupil tracking mode.
        The key idea is to estimate the systematic noise in a pixel by regressing the time series of this pixel onto a set of causally independent, signal-free predictor pixels.
        We use regularized linear models in this work; however, other (non-linear) models are also possible.
        In a second step, we demonstrate how the HSR framework allows us to incorporate observing conditions such as wind speed or air temperature as additional predictors.
    }{
        When we apply our method to four data sets from the VLT/NACO instrument, our algorithm provides a better false-positive fraction than PCA-based PSF subtraction, a popular baseline method in the field.
        Additionally, we find that the HSR-based method provides direct and accurate estimates for the contrast of the exoplanets without the need to insert artificial companions for calibration in the data sets.
        Finally, we present first evidence that using the observing conditions as additional predictors can improve the results.
    }{
        Our HSR-based method provides an alternative, flexible and promising approach to the challenge of modeling and subtracting the stellar PSF and systematic noise in exoplanet imaging data.
    }
    
    
    \maketitle

\section{Introduction}
\label{sec:introduction}

Since the first-ever image of a planet outside of our solar system was released not even twenty years ago \citep{Chauvin_2004}, direct imaging of extrasolar planets has come a long way.
Today, the detection and characterization of exoplanets through high-contrast imaging (HCI) is pursued at all major ground-based observatories, and several dedicated new instruments such as VLT/ERIS \citep{Davies_2018} and ELT/METIS \citep{Brandl_2016} will come online in the next few years seeking to push the limits of our observations from young hot gas giants to terrestrial planets.
However, high-performance instrumentation is only one of at least two ingredients required to make high-contrast imaging work.
Besides the hardware, we also require powerful algorithms for data post-processing that can separate the signals of a planet and its host star.
One key challenge here is the expected contrast: 
Even at favorable wavelengths in the near- and mid-infrared, stars outshine their companions by several orders of magnitude, with flux ratios ranging from $10^{-3}$--$10^{-4}$ for hot Jupiters down to $10^{-10}$ for Earth-like planets in the habitable zone (see, \eg \citealt{Traub_2010}).
Further challenges arise from the time-variability of the instrumental point spread function (PSF) (\eg due to the changing atmosphere and time-variable instrument performance) and from \emph{speckles}.
Speckles are a form of \enquote{structured noise, with both spatial and temporal correlations}~\citep{Males_2021} that occurs when scattered light from the star produces \enquote{transient spots that move about as mutually coherent patches of phase [...] happen to combine constructively in the image plane}~\citep{Bloemhof_2001}.
Based on their origin, one can distinguish between \emph{quasi-static speckles}, which are due to \enquote{imperfections within the telescope and instrument optics}, and \emph{atmospheric speckles}, which happen when atmospheric turbulence causes perturbations of the stellar wavefront~\citep{Males_2021}.
Since speckles often mimic the signal that we expect to see from an extrasolar planet, they are the major challenge in high-contrast imaging and \enquote{the limiting noise source in long exposure coronagraphic observations}~\citep{Males_2021}.

\paragraph{Current state of the art:}
\label{subsec:state-of-the-art}

In this work, we focus on post-processing HCI data obtained in pupil tracking mode, that is, with the field derotator of the instrument switched off.
For brevity, we also refer to this simply as angular differential imaging (ADI) \citep{Marois_2006}.
Furthermore, we limit ourselves to imaging \emph{point sources} (\ie exoplanets or brown dwarfs) and leave aside the growing body of work on imaging extended structures, such as circumstellar disks.
Even with this limited focus, there exists already a sizeable number of algorithms in the literature. 
Based on a taxonomy suggested by \citet{Cantalloube_2020}, we can distinguish between two main classes of methods:

\begin{description}
    \item[\textbf{Speckle subtraction techniques:}] 
        Algorithms in this category construct an explicit model of the stellar PSF, which is subsequently subtracted from the original data.
        The resulting residual is an estimate of the planet signal in units of flux.
        The most popular algorithm from this category is PCA-based PSF subtraction, also known as KLIP \citep{Soummer_2012, Amara_2012}.
        Other examples include LOCI \citep{Lafreniere_2007} and its variants TLOCI \citep{Marois_2014} and MLOCI \citep{Wahhaj_2015}, SNAP \citep{Thompson_2021}, NNMF \citep{Arcidiacono_2018} and LLSG \citep{GomezGonzalez_2016}.
    \item[\textbf{Inverse problem techniques:}] 
        Algorithms in this category model and track a planet signal using forward models.
        Their output consists of detection maps where the value of a pixel is not in units of flux but describes some form of probability that the pixel contains a planet.
        Examples from this category include ANDROMEDA \citep{Mugnier_2009,Cantalloube_2015}, FMMF \citep{Ruffio_2017}, PACO \citep{Flasseur_2018} and TRAP \citep{Samland_2021}.
\end{description}

Moreover, there is also a growing number of \enquote{meta techniques}, which are algorithms that take the output of another post-processing algorithm (typically from the speckle subtraction category) as their input and combine or process it further to improve the results.
Examples here include the RSM detection map introduced by \citet{Dahlqvist_2020, Dahlqvist_2021} and the SODINN/SODIRF algorithm proposed by \citet{GomezGonzalez_2018}.

Finally, several works have studied how to evaluate and quantify the results of a post-processing algorithm.
This includes, for example, estimating the statistical significance of a detection using a $t$-test \citep{Mawet_2014}, the computation of time domain detection maps \citep{Pairet_2019}, or assessing the results of a post-processing algorithm using performance maps \citep{JensenClem_2017}.
Nevertheless, computing robust detection limits that do not rely on assumptions about the distribution of the residual noise and meaningfully comparing different algorithms---especially across the two categories described above---currently remain challenging problems \citep[cf.][]{Bonse_inprep} which we deem beyond the scope of this work.

\paragraph{Contributions:}
\label{subsec:contributions}

We propose a new algorithm for post-processing HCI/ADI data sets that we derive from a specific motivation: the insight that there are some forms of prior knowledge about the data that existing algorithms do not fully appreciate.%
\footnote{
    An early version of this work was presented in the form of a workshop contribution in \citet{Gebhard_2020}.
}
Specifically, we are referring to (1) the expected structure of the stellar PSF, including speckles, and (2) the causal model for the underlying data-generating process.
In terms of the taxonomy introduced in \cref{subsec:state-of-the-art}, our method is a speckle subtraction technique.
Using four data sets from VLT/NACO \citep{Lenzen_2003,Rousset_2003}, we first demonstrate that our proposed method is competitive with PCA-based PSF subtraction.
Subsequently, we study selected properties of our algorithm in more detail and show, for example, that our algorithm is more resistant to self- or over-subtraction of the planet signal (a common side effect in PCA-based PSF subtraction) and allows us, therefore, to estimate the brightness of a planet directly with good precision.
Finally, we demonstrate the flexibility of our proposed approach by extending it to incorporate the observing conditions of a data set and providing evidence that this can improve the denoising performance.
To foster further research in this area and improve reproducibility, we release our pre-processed data sets along with this paper and make all of our code publicly available on GitHub.%
\footnote{
    \url{https://github.com/timothygebhard/hsr4hci}
}

\begin{figure*}
    \centering
    \includegraphics[width=184mm]{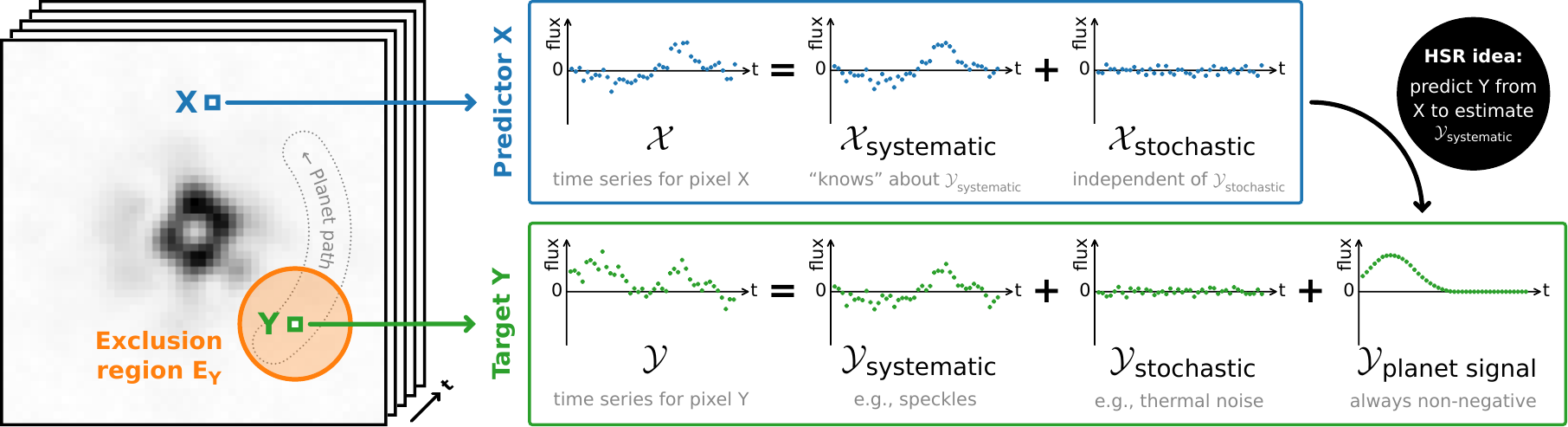}
    \caption{
        This figure shows the basic idea of using half-sibling regression for ADI data and introduces our notation.
        It also illustrates the causal model that we assume for the process that generates the data.
        \emph{Note:} In practice, we do not only use a single pixel $X$ as a predictor but a set of predictor pixels (see \cref{fig:predictor-selection}).
    }
    \label{fig:idea-and-notation}
\end{figure*}

\section{Prior domain knowledge about the problem}
\label{sec:scientific-domain-knowledge}

Even before we have made any actual observations, we have an idea of what we expect the data to look like in case they contain the signal from an exoplanet:

\begin{description}
    \item[\textbf{Planet signal:}]
        Depending on the type of coronagraph that is used during an observation, we know the (approximate) shape that the planet signal will have:
        For pupil plane coronagraphs, such as apodizing phase plate (APP) coronagraphs, the planet signal should match the shape of the stellar PSF, that is, basically an imperfect Airy pattern with additional instrumental effects such as the signature from the spiders that hold the secondary mirror in place.%
        \footnote{
            This does not hold for focal plane (\eg vortex) coronagraphs.
        }
        Additionally, we know that the signal from a planet is relatively sparse: even in the presence of one (or multiple) planets, only a small fraction of pixels on the detector will be affected.
    \item[\textbf{Temporal dynamics:}] 
        We know that in angular differential imaging, the rotation of the Earth causes the signal from a planet to describe a circular arc around the star.
        Of course, the exact starting position of this arc is unknown; however, we do know the opening angle of the arc (as it is given by the duration of the observation), as well as the angular velocity of the signal at any time.
        This, in turn, means that if we know the position of the planet at any given point in time $t$, we know its position at every point in time.
        Furthermore, we also know that speckles should not exhibit this temporal dynamic: they should either remain statically at the same position or move randomly (like the atmospherical perturbations that create them).
        In fact, this \emph{diversity} (\ie an aspect of the data for which planets and speckles are known to behave differently) is the main idea behind angular differential imaging.
\end{description}

Existing post-processing algorithms are already using these types of domain knowledge:
For example, the fact that the planet is not at the same position in all frames is why it may be assumed not to show up in the first $k$ principal components of the data, meaning that a projection onto those components gives an estimate of the systematic noise.
This is the key idea behind PCA-based PSF subtraction / KLIP \citep{Soummer_2012, Amara_2012}.
There is, however, even more domain knowledge that the existing literature does not yet fully appreciate:

\begin{description}
    \item[\textbf{Temporal order of the data:}]
        We know the \enquote{arrow of time} in our data. 
        However, for some algorithms (\eg PCA), the order of the frames is irrelevant: 
        You can randomly shuffle the data before you post-process it without changing the final result.
        Furthermore, PCA only looks at the variance of the data on a global time scale. 
        To distinguish speckles from planets, it might be helpful to look also at effects on short time scales, though.
        After all, the processes that affect the data generation are continuous (\eg air moving in the atmosphere), and we thus might expect continuity on short time scales.
    \item[\textbf{Expected structure of the PSF:}] 
        There exists an extensive body of theoretical work on the expected structure of the (stellar) PSF, including the structure of the speckle pattern; see, for example, \citet{Bloemhof_2001, Bloemhof_2002a, Bloemhof_2002b, Bloemhof_2003, Bloemhof_2004a, Bloemhof_2004b, Bloemhof_2004c, Bloemhof_2004d, Bloemhof_2006, Bloemhof_2007, Boccaletti_2002, Sivaramakrishnan_2002, Perrin_2003, Ribak_2008}.
        Two of the key findings from these results are:
        \begin{enumerate*}
            \item \emph{Speckle pinning}, that is, the insight that speckles are most likely to occur at (and be pinned to) the locations of the secondary maxima of the Airy pattern of the stellar PSF, and
            \item \emph{Speckle symmetry}, that is, a theoretical explanation why the speckle pattern should exhibit some degree of (anti-)symmetry across the center (\ie the location of the star).
        \end{enumerate*}
        Particularly the second result seems promising from the perspective of HCI post-processing:
        If we can assume speckles to be at least partially (anti-)symmetric, this could help to distinguish them from planet signals which should exhibit no such symmetries.
        \citet{Perrin_2003} even explicitly suggest that \enquote{[k]nowledge of this antisymmetry can be used to improve ways of obtaining and reducing high Strehl ratio imaging data} and recommend that \enquote{[w]hen reducing data, the antisymmetry of the first-order speckles is prior information that should be fed into deconvolution approaches.}
        Several works---for example, \citet{Boccaletti_2002}, \citet{Bloemhof_2007} and \citet{Dou_2015}---have, therefore, already suggested to post-process HCI data by subtracting from each frame a copy of the image that has been rotated by \SI{180}{\degree} (which, of course, only works in the case of symmetry, not antisymmetry).
        
        In \cref{sec:symmetries-in-hci-data-in-practice} we present three experiments in which we find clear empirical evidence for (anti)-symmetric structures in ADI data, which motivates us to pursue this direction further.
    \item[\textbf{Causal structure of the data-generating process:}] 
        The value of a pixel on the detector can generally be thought of as a sum of three terms (see \cref{fig:idea-and-notation}): 
        The planet signal (which may be $0$), the systematic noise (\eg speckles), and the stochastic noise (\eg photon noise from thermal background). 
        We might even be able to make reasonable assumptions about the respective distributions or potential correlations.
        Additionally, we have access to a set of observables that we can assume to have a causal effect on the data (in particular the systematic noise), namely the observing conditions.
\end{description}

In this paper, we develop an algorithm that focuses on incorporating the last two pieces of domain knowledge into a machine learning methodology, and we advocate for this as a potential direction for future research.
In this sense, we see our work in line with, for example, \citet{Ansdell_2018} who have argued for \enquote{the importance of including domain knowledge in even state-of-the-art machine learning models when applying them to scientific research problems that seek to identify weak signals in noisy data.}

We acknowledge that the TRAP algorithm \citep{Samland_2021}, which was developed independently and partly in parallel to this work, also uses several of these ideas.
It is similar to our method in the sense that it also takes its inspiration from the half-sibling regression framework of \citet{Schoelkopf_2016} and constructs a causal, temporal, regularized linear model for each pixel to fit the systematic noise.
More precisely, TRAP performs a joint fit of the systematics and the signal (by including a potential signal as a predictor), which is similar to the \enquote{signal fitting}-variant of our algorithm (see below).

A crucial difference between TRAP and our method is that we propose a speckle subtraction technique, whereas TRAP falls into the category of inverse problem techniques (cf. \citealt{Cantalloube_2020}).
This means we construct an explicit estimate for the systematic noise, which we subtract from the data to produce a residual. In contrast, TRAP uses its planet model to compute a detection map.
Another difference is the form of regularization: 
TRAP uses principal component regression, with a single global hyper-parameter controlling the regularization strength for all pixels, whereas we use ridge regression with a regularization strength that is determined automatically for each pixel via efficient leave-one-out cross-validation.%
\footnote{
    As Sect. 1.3.1 of \citet{VanWieringen_2015} points out, there is a connection between principal component regression and ridge regression: 
    Principal component regression can be seen as thresholding the singular values of the design matrix $\Xts$ (\ie a discrete map), whereas ridge regression corresponds to shrinking them (\ie a continuous map).
}
Other differences include our usage of a $k$-fold cross-validation scheme during training, the exact choice of the predictor set and the exclusion region, as well as the selection of the final residuals (our \enquote{stage 2}; see below).
Finally, in this work, we also study the use of metadata in the form of observing conditions as additional predictors, which \citet{Samland_2021} only mention as potential future work.

\section{Method}
\label{sec:method}

Our proposed algorithm consists of a modified version of half-sibling regression (HSR), the statistical learning algorithm underlying the causal pixel model (CPM) approach of \citet{Wang_2016, Wang_2017}.
Half-sibling regression \citep{Schoelkopf_2016} is a conceptually simple yet flexible denoising technique that, at its core, is based on the assumption of a particular causal model for the data-generating process.
It was originally proposed to process data of the \emph{Kepler} mission, but has also been applied in other domains; for example, remote sensing \citep{Kondmann_2021}.

In this section, we first explain the general idea of HSR before discussing the specific modifications that we propose to apply the method to ADI data.
For a more technical explanation of half-sibling regression that closely follows the descriptions in \citet{Schoelkopf_2016}, we refer the reader to \cref{sec:hsr-in-detail}.

\subsection{Half-sibling regression: the general idea}

Assume we are looking at a pixel \Ypos of the detector of a telescope.
As discussed before, our understanding of the causal structure of the data-generating process suggests that the value of this pixel should consist of three terms (see also \cref{fig:idea-and-notation}): 
\begin{enumerate*}
    \item the signal from an exoplanet (which may be $0$), 
    \item the systematic noise, and
    \item the stochastic noise.
\end{enumerate*}
Our goal with half-sibling regression is to remove the systematic noise component from \Ypos to recover an approximation of the planet signal.
To this end, we look at another pixel \Xpos.
If we choose \Xpos such that the distance between \Xpos and \Ypos is greater than the expected diameter of a planet signal, which is known a priori (typically a few pixels), we can safely assume that \Xpos and \Ypos are not affected by the same planet signal.
Using the time series \Yts for \Ypos and \Xts for \Xpos, we then build a regression model to predict the value of \Ypos from the value of \Xpos (\ie we learn an estimate for $\mathbb{E}(\Ypos\,|\,\Xpos)$).
Because \Xpos does not know anything about the planet signal at \Ypos, the prediction should not contain information about that signal.
However, since the pixels \Xpos and \Ypos are recorded with the same instrument, their systematic noise components share some mutual information. 
This means that \Xpos should be able to predict---at least partially---the systematic noise at \Ypos.
We denote this prediction, which is also a time series, as $\hat{\Yts}$.
If we now subtract $\hat{\Yts}$ from the original \Yts, we get a residual time series $\mathcal{R} = \Yts - \hat{\Yts}$ which still contains the planet signal (and stochastic noise) in \Yts but no longer the systematic noise.

Note how it is crucial that \Xpos must not know anything about the planet signal in \Ypos.
If this is not the case, the prediction based on \Xts will contain (part of) the signal in \Yts. 
If we then subtract $\hat{\Yts}$ from the true \Yts, we remove (part of) what we are trying to find.
The choice of \Xpos will, therefore, always depend on \Ypos.

In practice, we do not have to limit ourselves to a single predictor pixel \Xpos.
Generally speaking, we can improve the prediction of the systematic noise by choosing a whole set of pixels as long as all pixels in the set are \emph{causally independent} from \Ypos.
Our domain knowledge about the problem can guide the search for a suitable set of predictor pixels; we discuss this in detail below.

\vspace{0.5cm}
{
    \itshape
    \footnotesize
    \textbf{Notation:} 
    In the remaining part of this work, we will use $T$ to denote the number of time steps, or frames, in an observation.
    We further use \Ypos to denote a target position (\ie a pixel) on the detector and \Yts for the corresponding time series, which is a vector in $\mathbb{R}^{T}$.
    Furthermore, for a given target position \Ypos, $\Xpos = \Xpos(\Ypos)$ denotes an ordered set of $d$ positions (\ie pixels) that we use as predictors for \Ypos. 
    Finally, $\Xts \in \mathbb{R}^{T\,\times\,d}$ refers to the matrix whose columns are the time series for positions in \Xpos.
}

\subsection{Applying HSR to Angular Differential Imaging}
\label{subsec:applying-hsr-to-adi}

In this part, we discuss the steps and various subtleties that are required to translate the intuition explained in the last section into an actual denoising algorithm for (ADI-based) HCI data.
We begin with the most basic version, which we dub \enquote{vanilla half-sibling regression}.
Given an ADI data set, it works as follows:
\begin{enumerate}
    \item Choose a region of interest (ROI) around the star.
    \item Loop over all pixels in the ROI.
        For each pixel \Ypos in the ROI, do the following (each step is discussed in more detail below): 
    \begin{enumerate}
        \item Determine an \emph{exclusion region} $E(\Ypos)$, that is, a set of pixels which are too close to \Ypos to be causally independent and must, therefore, not be used as predictors (see above).
        \item Excluding $E(\Ypos)$, choose a set \Xpos of $d$ predictor pixels that are causally independent of a potential planet signal at \Ypos.
        \item Learn a model $m$ by regressing \Yts onto \Xts.
        \item Get the model prediction $\hat{\Yts} = m\left( \Xts \right)$ and compute the corresponding residual time series $\mathcal{R}(\Ypos) = \Yts - \hat{\Yts}$.
    \end{enumerate}
    \item Once we have one residual time series $\mathcal{R}(\Ypos)$ for all \Ypos in the ROI, derotate each frame in the resulting residual stack by its respective parallactic angle and compute the mean of the derotated stack along the temporal axis.
    We call the resulting frame the \emph{signal estimate} for the data set.
\end{enumerate}

Let us now take a closer look at the sub-steps of step 2.
All of the following applies both to the vanilla version of HSR and also to the more advanced version that we describe in \cref{subsec:two-stage-hsr}.

\subsubsection{Choosing an exclusion region and a set of predictors}
\label{subsubsec:predictor-selection}

\begin{figure}[t]
    \centering
    \includegraphics[width=90mm]{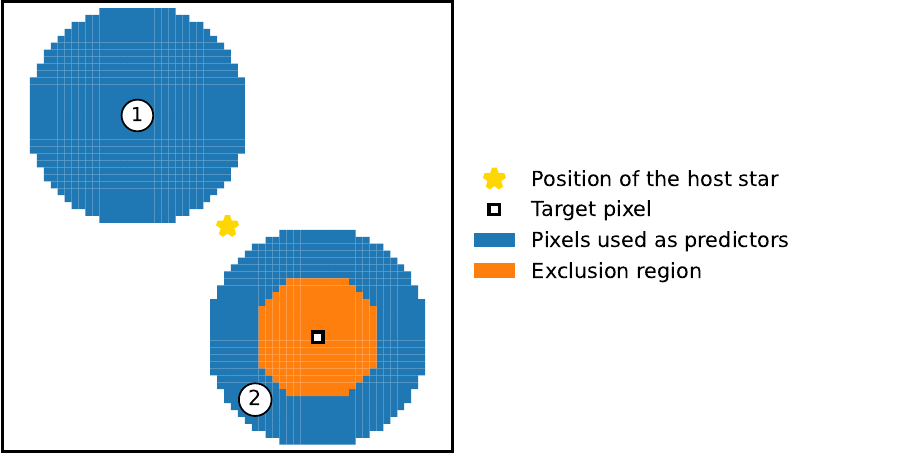}
    \caption{
        This figure illustrates our choice of predictors pixels $\Xpos$ and the exclusion region $E(\Ypos)$ for learning a model to predict the target pixel $\Ypos$.
        For the motivation and details, see \cref{subsubsec:predictor-selection}.~%
        \href{https://github.com/timothygebhard/hsr4hci/tree/master/scripts/figures/predictor-selection/make_plot.py}{\LinkToCode}
    }
    \label{fig:predictor-selection}
\end{figure}

The exclusion region $E(\Ypos)$ for a pixel \Ypos is the set of pixels that contain information about the presence of a planet signal at \Ypos, that is, pixels that are not causally independent of such a signal.
One simple choice for $E(\Ypos)$  is to exclude all pixels inside a given radius around $\Ypos$.
For this work, we have chosen a radius of 9 pixels, which equals approximately 2 FWHM of the PSF.
A more sophisticated approach for determining $E(\Ypos)$ could also consider the shape of the PSF and the planet's movement over time.

When it comes to choosing a set of predictor pixels $\Xpos(\Ypos)$, we could, in principle, use all pixels that are not part of the exclusion region.
However, using so many predictors (\ie potentially tens of thousands) is computationally expensive. 
Models with many predictors also require many learnable parameters, making them more susceptible to overfitting (see below).
Finally, we already know that not all pixels are equally informative, and we expect some pixels to \enquote{know more} about the systematic noise in \Ypos than others.
Based on our domain knowledge, we, therefore, propose to use the following set of pixels as predictors:
First, a circular region symmetrically across the origin from \Ypos, because this region should ideally contain information about whether or not \Ypos contains a speckle (see \cref{sec:symmetries-in-hci-data-in-practice} where we study the existing symmetries in HCI data sets in practice).
Second, a region around $Y$ itself, to capture any \enquote{local} effects.
We show an example of such a choice of predictors, including the corresponding exclusion region, in \cref{fig:predictor-selection}.
In this work, we set the radius of the two predictor regions to 16 pixels; no optimization was performed here.

This choice of predictors and exclusion region constitutes a simplification of the respective choices presented in our preliminary version of this work \citep{Gebhard_2020}.
Experimentally, we have not found a significant difference between the two and have, therefore, decided to proceed with the simpler option.

In practice, we do not have to limit our choice of predictors for \Ypos only to other pixels from the same detector:
We can also choose other quantities as predictors that we assume to be informative about the systematic noise in \Ypos.
In particular, we can also use quantities that are a \emph{cause} of the noise, not an \emph{effect}.
\citet{Schoelkopf_2016} justify this \enquote{prediction based on non-effects of the noise variable} by arguing that the direction of the causal relationship between the predictors $X$ and the systematic noise $N$ does not matter for the idea of half-sibling regression.
In our specific problem setting, an intuitive choice could be to incorporate the observing conditions into our model: 
After all, we know that parameters like the seeing, wind speed, or coherence time should be related to the systematic noise at \Ypos and could, therefore, potentially be helpful to estimate this systematic noise.
Moreover, these quantities are causally independent of a potential planet signal at \Ypos.
We study this idea in more detail in \cref{sec:observing-conditions}.

Finally, \citet{Schoelkopf_2016} also suggest that one can regress \Ypos onto its \emph{own} past and future if one can assume a signal with compact (temporal) support.
While this may have worked well for the case of transit photometry, where the signal---that is, the dips in the light curves---is short (\eg a few hours) compared to the duration of the observation (several weeks or months), this would be much harder in the case of ADI data:
First, for planets close to the star or data sets with a low field rotation, a planet might be present in a given pixel for a large fraction of the observation.
Second, due to frame selection and the fact that the data is typically recorded in cubes, the time between two consecutive frames in a data set is usually not always the same.
For the present work, we have therefore decided not to investigate this idea further.

\subsubsection{Learning the model(s) for a pixel}
\label{subsubsec:learning-a-model}

In order to predict the value of a pixel \Ypos from a set of predictors \Xpos, we need to learn a model $m$.
The half-sibling regression framework is very flexible in this regard: 
In principle, we can use \emph{any} machine learning model for regression, from simple linear models or random forests to kernel methods or neural networks.
Generally, more powerful models will be able to capture more complex relations between the predictors and the targets.
However, more powerful models also require more training data, have more hyper-parameters that require tuning, and are more susceptible to overfitting.
Also, they usually take more time to train; an effect that may quickly add up, considering we have to train at least one model per pixel in the ROI.
Therefore, in this work, we limit ourselves to regularized linear models, specifically ridge regression (see, \eg Sect. 3.4 of \citealt{Hastie_2009}).

As ridge regression is sensitive to scale differences in the data, we start out by normalizing all predictors using a $z$-transform, that is, we subtract from each column of $\Xts$ its mean and divide by its standard deviation.
We then train the model $m: \mathbb{R}^d \to \mathbb{R}$ with parameters $\bm{\beta}$ (a vector whose dimensionality depends on the exact choice of model) by minimizing the sum of a loss function $\mathcal{L}: \mathbb{R} \times \mathbb{R} \to \mathbb{R}$ over all $T$ frames:
\begin{equation}
    \min_{\bm{\beta}} \sum_{t=1}^{T} \mathcal{L}\left( \Yts_t, \hat{\Yts}_t \right) 
    = \min_{\bm{\beta}} \ \sum_{t=1}^{T} \mathcal{L}\left( \Yts_t, m(\Xts_t\,;\,\bm{\beta}) \right) \,.
    \label{eq:minimization-objective}
\end{equation}
The loss function $\mathcal{L}$ measures the difference between the target, $\Yts_t$, and model prediction $\hat{\Yts}_t$.
A common choice for $\mathcal{L}$ is the square of the prediction error, $(\Yts_t - \hat{\Yts}_t)^2$, plus some regularization term that depends on $\bm{\beta}$.
For the specific case of ridge regression, where $\bm{\beta} \in \mathbb{R}^d$ and $m(\Xts_t\,;\,\bm{\beta}) = \Xts_t\,\bm{\beta}$, the objective is given by:
\begin{equation*}
    \min_{\bm{\beta}} \ \left\Vert \Yts - \Xts \bm{\beta} \right\Vert^2_2 + \lambda \cdot \left\Vert \bm{\beta} \right\Vert^2_2 \,,
\end{equation*}
where $\lambda > 0$ is the regularization strength (\ie a hyper-parameter).
This is a particularly convenient choice for the minimization objective because it has the following analytical solution:
\begin{equation}
    \bm{\beta} = (\Xts^\top \Xts + \lambda \cdot I_d)^{-1} \Xts^\top \Yts \,.
\end{equation}
Additionally, the value of $\lambda$ can be chosen automatically, for example using computationally efficient leave-one-out cross-validation (see, \eg Sect. 1.8.2 of \citealt{VanWieringen_2015} for details).

To further reduce the risk of overfitting, we use a particular form of $k$-fold cross-validation (see, \eg Sect. 7.10 of \citealt{Hastie_2009}):
We do not learn just one model per pixel but instead, we split the target and predictor time series into $k$ sub-time series.
For each of these sub-times series, we then train a model on the respective other $k - 1$ sub-times series and compute the residual by applying the model to the held-out data to compute the residual.
To formalize this procedure, let:
\begin{equation}
    S = \lbrace 1, 2, \ldots, T \rbrace \,
\end{equation}
be the set of temporal indices for our data set.
We now split $S$ into $k$ disjoint subsets that we define as follows:
\begin{equation}
    S_{i = 1, \ldots, k} 
    = \lbrace t \in S \ |\ t \equiv i - 1\ \text{mod}\ k \rbrace 
    = \lbrace i, i + k, i + 2k, \ldots \rbrace
    \,. 
\end{equation}
This particular splitting scheme has the advantage that the field rotation in each split matches the field rotation of the full data set.
For each index set $S_i$, we then learn a model~$m_i$ with parameters~$\bm{\beta}_i$ by minimizing \cref{eq:minimization-objective} for $t \in S\,\setminus\,S_i$ instead of $t \in S$.
The final residual time series for a pixel is then obtained as:
\begin{equation}
    \mathcal{R} 
    = \mathcal{R}[S] = \bigcup_{i=1,\ldots,k} \mathcal{R}[S_i]%
    \,,\enspace\text{for}\enspace%
    \mathcal{R}[S_i] := \Yts[S_i] - m_i(\Xts[S_i]) \,.
\end{equation}
Generally speaking, increasing $k$ will usually improve the results (as we train each model on more data); however, this also drives up the computational cost.
For this work, we have chosen $k = 3$.

\subsection{Two-stage half-sibling regression}
\label{subsec:two-stage-hsr}

\subsubsection{Practical challenges for vanilla HSR}
In principle, the vanilla version of HSR described in the last section is already a functioning denoising algorithm for ADI data.
In practice, however, its performance is often not competitive as we find that the HSR seems to remove substantial parts of the planet signal, especially for bright planets.
Looking into this effect in more detail, we found that this has at least two reasons:

\paragraph{Overfitting:}
Even if we carefully choose the predictors \Xpos such that they are \emph{causally independent} from the target \Ypos, we may still be able to (accidentally) model the signal component in \Ypos due to \emph{overfitting}.
Overfitting means that the model does not learn to capture the \enquote{true} relationship between \Xpos and \Ypos, but instead learns a mapping that only holds on the training data.
For example, by (partially) memorizing the training data, the model can fit features of \Yts even if \Xts does not actually contain any information about those features. 
In this case, our prediction for \Yts can accidentally contain some part of the signal we are trying to preserve, which is subsequently lost when we compute the residual.
Overfitting can happen even for simple linear models, especially if the number of predictor pixels is large compared to the number of frames.

\begin{figure*}
    \centering
    \includegraphics[width=184mm]{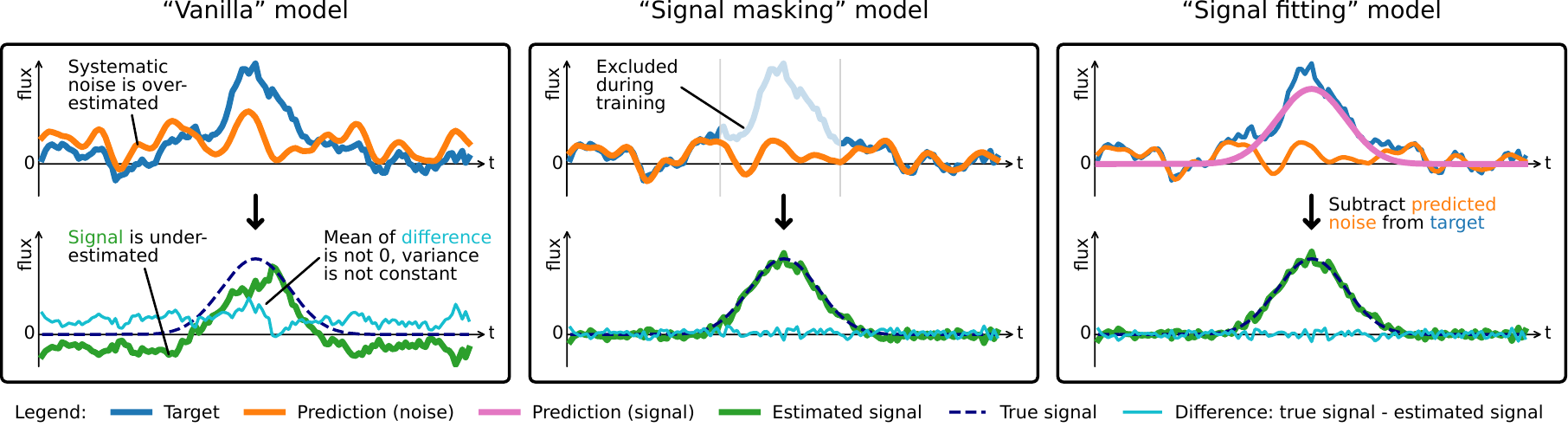}
    \caption{
        This figure illustrates the effect of the expected value of the planet signal on vanilla HSR for a target pixel that contains a planet exactly at half time of the observation and compares it with \emph{signal masking} and \emph{signal fitting}.
        For the sake of clarity, and to separate the effect from the impact of overfitting, we work with a toy model where the target time series is a weighted sum of an artificial planet signal, three predictor time series (the sum of which we call the systematic noise), and another time series consisting solely of white noise (the stochastic noise).
        We train a standard linear regression model (ordinary least squares) by regressing the target time series onto the three predictor time series.
        For the vanilla HSR model, we use the entire target time series for training.
        For the \emph{signal masking} model, we exclude all time steps at which the true signal time series exceeds \SI{20}{\percent} of its maximum (indicated by the shaded region).
        Finally, for the \emph{signal fitting} model, we add the true signal time series as an additional predictor to the model.
        We then apply the trained model to the entire predictor time series to predict the noise (shown in orange).
        For the signal fitting model, we have to set the coefficient corresponding to the predictor containing the true signal to 0 to get the \enquote{noise only} prediction.
        Once we have the prediction for the noise, we subtract it from the target to get our estimate for the signal, shown in green on the bottom row.
        We find an excellent agreement between the estimated and the true signal for the signal masking and signal fitting models.
        For the vanilla model, we observe that the presence of the planet signal has two effects which we describe in the main text.
    }
    \label{fig:expected-valued}
\end{figure*}

\paragraph{Self-subtraction:}
Besides overfitting, there is also the problem that the expected value of a planet signal in a pixel---that is, its average over time---is not zero.
This is problematic since, as described above, we learn the HSR model for a given target pixel by minimizing the mean squared difference (along the temporal dimension) between the observed value pixel value and the model prediction.
If the target time series now contains a \enquote{planet bump}, this bump cannot be modeled by the predictor time series (because the predictors were chosen to be causally independent of the target).
However, the bump still affects the fit of the model in two ways, which we also illustrate in \cref{fig:expected-valued}.

First, it causes the model to learn a constant offset that approximately matches the average value of the signal time series (\ie the expected value).
As a result, the model overestimates the systematic noise by this amount.
At the times where the target time series contains planet signal, this then turns into self-subtraction, where we lose part of the planet signal due to the over-estimated systematic noise.
This effect gets worse the brighter the planet is because a stronger planet signal will impact the fit more than a faint one.

Second, as the model tries to explain the signal bump using only predictors that do not contain such a bump, it usually ends up choosing a combination of predictors that explains the systematic noise at points where the signal is 0 worse than if we had ignored the time steps with a non-zero signal during training.

\subsubsection{Possible remedies: masking or fitting the signal}
\label{subsec:fitting-or-masking}

To address the challenges presented in the last section, we propose two potential modifications of the vanilla version of half-sibling regression: \emph{signal masking} and \emph{signal fitting}.

\paragraph{Signal masking:}
The idea of \emph{signal masking} is the following:
Assume we knew at which time the planet signal reaches its peak in a given pixel.
In this case, we could use the parallactic angles to compute the expected planet signal time series $\Yts_\text{ps}$ for this pixel.%
\footnote{
    In our current implementation, we use the unsaturated PSF template that is usually recorded for HCI data sets to compute the expected signal time series.
    Of course, in doing so, we are making several simplifying assumptions: 
    For example, we treat the instrumental PSF as temporally and spatially constant and ignore the influence of the coronagraph.
    We leave it to future research to investigate if the performance of our method can be improved through a more sophisticated approach to compute $\Yts_\text{ps}$.
}
Using this expected signal, we could then determine a mask to exclude the time steps at which the planet flux exceeds a certain threshold from the training data for the noise model.
This means we would effectively only use the frames in which no planet is present in the target pixel to learn the noise model.
Formally, we would find a set of time steps $S_\text{ps} \subset S$:
\begin{equation}
    S_\text{ps}
    = \lbrace t \in S \:|\: \Yts_\text{ps} > \text{threshold} \rbrace \,,
\end{equation}
for some given threshold.
Now, when we learn a model~$m_i$ (see \cref{subsubsec:learning-a-model}), we minimize over $S \setminus (S_i \cup S_\text{ps})$ instead of $S \setminus S_i$.

Of course, in practice, we do not a priori know at which time---if at all---a given pixel contains a planet.
To address this, we add another loop to the procedure outlined in \cref{subsec:applying-hsr-to-adi}:
For every pixel \Ypos in the ROI, we loop over a temporal grid of possible times~$t$.
As the masking does not need to be frame-perfect, the size of the temporal grid can be significantly smaller than the total number of frames.
Then, for every such combination of \Ypos and $t$, we compute $S_\text{ps}$ and train the corresponding models as suggested above.
Once we have trained a model for each potential time $t$, we need to choose which of them we want to use to denoise \Yts.
We describe our approach for this in \cref{subsubsec:two-stage-hsr}.

Besides increasing the number of models that we need to train, one downside of this masking approach is that we cannot use all of our data for training:
Masking out frames for the training data is similar to the frame exclusion required by algorithms such as LOCI.
Especially at small separations or for small field rotations, we may find that we have to exclude many frames and train only on a small fraction of our data.

\paragraph{Signal fitting:}
The basic idea of signal fitting is very similar to signal masking.
Again, for each spatial position \Ypos, we loop over a temporal grid of times $t$ at which the planet signal in \Ypos reaches its peak and compute the corresponding $\Yts_\text{ps}$.
However, instead of using $\Yts_\text{ps}$ to determine the time steps that we exclude from training, we use $\Yts_\text{ps}$ as an additional predictor for our models.
In this case, we learn to simultaneously predict both the systematic noise and the signal (under the current hypothesis for the planet path).

This implies certain requirements for the model we are learning.
Because we assume the systematic noise and the signal to interact additively, our model $m$ should have the following structure:
\begin{align}
    m\left( \Xts, \Yts_\text{ps} \right) = 
    m_\text{noise}( \Xts ) + 
    m_\text{signal}\left( \Yts_\text{ps} \right) \,.
\end{align}
Since the photon flux from a planet cannot be negative, we would like to require the output of $m_\text{signal}$ to be strictly non-negative.
Furthermore, we need to be able to access $m_\text{noise}$ and $m_\text{signal}$ separately because ultimately, we only want to subtract the prediction of the systematics model from the data to estimate the signal.%
\footnote{
    Alternatively, one could deviate from the original HSR idea, discard the prediction from $m_\text{noise}$ and work with $m_\text{signal}$ instead.
    This is basically the approach of the TRAP algorithm from \citet{Samland_2021}.
    Note that in this case, the resulting algorithm is no longer a speckle subtraction technique (see \cref{subsec:state-of-the-art}).
    Another thing to consider when only keeping the output of $m_\text{signal}$ is how to deal with the part of the data that was not explained by the fit, that is, the difference between the original data and the sum of the signal and the noise model.
    In our method, we keep the unexplained part of the data as residual noise in the signal estimate.
}
For a linear model, the additivity and individual accessibility of $m_\text{noise}$ and $m_\text{signal}$ is trivially fulfilled.
Likewise, one could also construct a neural network architecture consisting of two independent sub-networks (for the noise and the signal) whose outputs are only summed up in the final layer.
However, for other non-linear regression methods, satisfying these constraints is much more challenging, which limits the flexibility of the HSR framework.

Finally, we need to balance $m_\text{signal}$ and $m_\text{noise}$ in the sense that if both model parts can explain the same aspect of the data, we need to decide which model should take precedence.
In the case of ridge regression, we can control the trade-off between  $m_\text{signal}$ and $m_\text{noise}$ by rescaling the respective predictors, in a way akin to the \emph{feature weighting} from \citet{Hogg_2021}:
By choosing $\Yts_\text{ps}$ on a larger scale than \Xts (which we can achieve by multiplying with a large number), we prioritize $m_\text{signal}$, as using the signal component of $m$ is now cheaper than the noise component in terms of the loss that is incurred through the $\ell_2$-penalty on the coefficients.
A potential downside of this approach is that it introduces an additional hyperparameter.
In our implementation, we multiply $\Yts_\text{ps}$ with a factor of $\num{1000}$ before using it as a predictor. 
This value was an ad-hoc choice that we did not optimize.

Another potential problem with signal fitting arises at small separations from the star or for small field rotations:
In these cases, $\Yts_\text{ps}$ can take on a rather generic form (instead of a clear \enquote{bump}-shape), and as a result, $m_\text{signal}$ may start to fit arbitrary trends in the data that should be part of the noise model.

\subsubsection{Putting it all together: two-stage half-sibling regression}
\label{subsubsec:two-stage-hsr}

\begin{figure*}[t]
    \centering
    \begin{minipage}[t]{184mm}
    \begin{subfigure}[t]{34mm}
        \centering
        \includegraphics[scale=1]{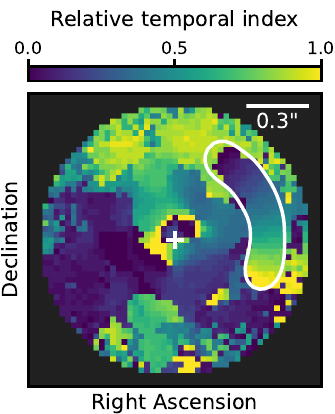}
        \subcaption{}
        \label{fig:hypothesis-map}
    \end{subfigure}%
    \hfill%
    \begin{subfigure}[t]{34mm}
        \centering
        \includegraphics[scale=1]{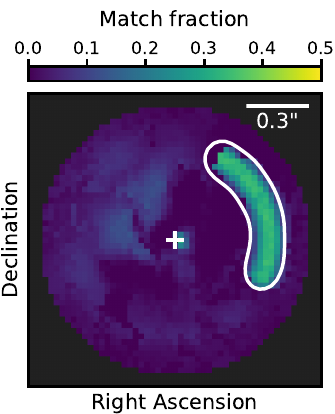}
        \subcaption{}
        \label{fig:match-fraction-map}
    \end{subfigure}%
    \hfill%
    \begin{subfigure}[t]{34mm}
        \centering
        \includegraphics[scale=1]{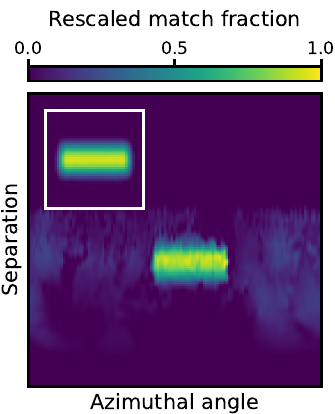}
        \subcaption{}
        \label{fig:polar-match-fraction-map}
    \end{subfigure}%
    \hfill%
    \begin{subfigure}[t]{34mm}
        \centering
        \includegraphics[scale=1]{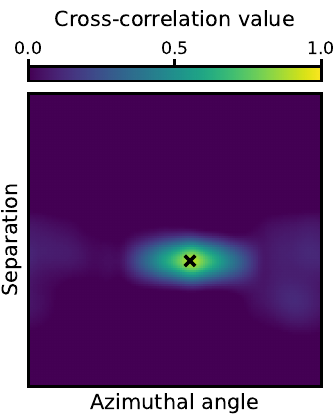}
        \subcaption{}
        \label{fig:cross-correlation-and-template}
    \end{subfigure}%
    \hfill%
    \begin{subfigure}[t]{34mm}
        \centering
        \includegraphics[scale=1]{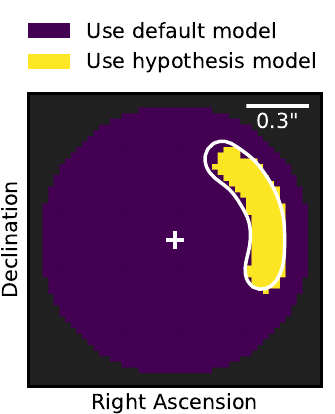}
        \subcaption{}
        \label{fig:residual-selection-mask}
    \end{subfigure}%
    \end{minipage}
    \caption{
        Outputs of the different steps of stage 2:
        (a) hypothesis map,
        (b) match fraction map,
        (c) polar match fraction map with expected signal template (in upper left corner),
        (d) cross-correlation between the polar match fraction match and the expected signal template, and
        (e) residual selection mask.
        The white lines in (a), (b), and (e) indicate the true trajectory of the planet in the data.
        We notice several things: 
        In (a), the pixels on the true planet trajectory show a clear gradient from early to late that matches the planet's movement.
        In (b), we then find that only these pixels produce a high match fraction, meaning that only their hypotheses are consistent with the rest of the data.
        Finally, in (e), we see a residual selection mask that matches the true planet trajectory almost perfectly.~%
        \href{https://github.com/timothygebhard/hsr4hci/blob/master/scripts/experiments/evaluate-and-plot/plot_results_of_stage_2.py}{\LinkToCode}
    }
    \label{fig:stage-2}
\end{figure*}

Both signal masking and signal fitting use a loop over possible times $t$ at which a given pixel contains a planet, meaning that after training, we have $n_\text{signal times} \leq T$ residual time series per pixel. 
As mentioned above, we would usually choose $n_\text{signal times} \ll T$.
Additionally, we also have a residual time series for the hypothesis that the pixel does not contain a planet which we obtain using the vanilla version of the method. 
Thus, the total number of residual time series per pixel is $n_\text{signal times} + 1$.
For each pixel, we now need to decide which of these residual time series we want to use in the construction of the final residual stack.
To this end, we introduce a new residual selection step, thus turning the half-sibling regression approach into a two-stage process.
The first stage consists of training the models for each pixel and computing the $n_\text{signal times} + 1$ residual time series, as described in the previous sections.
Stage 2 also consists of multiple sub-steps, which we summarize in \cref{alg:overview-stage-2} and illustrate in \cref{fig:stage-2}.
To not make the following descriptions too technical, we restrict ourselves to a high-level outline here and invite the interested reader to take a look at our code for further details.

\begin{algorithm}[t]
	\SetArgSty{}
	\DontPrintSemicolon
	\setlength{\parskip}{2.3pt}
	{
        Initialize empty hypothesis map (HM) \;
        Initialize empty match fraction map (MFM) \;
    	\ForEach{pixel $Y$ in the ROI}{
    	    $t^*$ $\leftarrow$ compute hypothesis time for $Y$ \;
    		MF $\leftarrow$ compute match fraction for $(Y, t^*)$ \;
    		Store $t^*$, MF at position $Y$ in HM, MFM \;
    	}
    	Project MFM to polar coordinates \;
    	Cross-correlate polar MFM with expected signal \;
    	Use blob finder to get peaks in cross-correlation result \;
    	Construct residual selection mask from peaks \;
        Use selection mask to select residual time series \;
    	Derotate residual stack (RS) using parallactic angles \;
    	Temporally average derotated RS to get signal estimate \;
	}
	\caption{Overview of stage 2}
	\label{alg:overview-stage-2}
\end{algorithm}

We begin stage 2 with the computation of a \emph{hypothesis map}.
Consider a pixel \Ypos: 
For each of the $n_\text{signal times}$ residual time series $\mathcal{R}(\Ypos, t)$, we compute the cosine similarity (CS) with the respective planet signal time series $\Yts_\text{ps}(t)$.
The CS takes on values in $[-1, 1]$, where $1$ means that the two time series are identical (up to a constant scaling factor) and $0$ means they are orthogonal.
We call the time $t^*$ that yields the highest CS value the \emph{hypothesis time} for \Ypos; see \cref{alg:hypothesis-time} for more details about how we compute $t^*$.
The interpretation of $t^*$ is basically: \enquote{If \Ypos contains a planet, then we believe that the peak of the planet signal occurs at $t^*$}.
We can create a 2D array in which every position \Ypos contains its respective candidate time $t^*(\Ypos)$ and call this our \enquote{hypothesis map}.

\begin{algorithm}[t]
	\SetArgSty{}
	\DontPrintSemicolon
	\setlength{\parskip}{3.1pt}
	{
    	$t^*$ $\leftarrow$ NaN\;
    	maximum $\leftarrow$ 0\;
    	\ForEach{time $t$ in temporal grid of size $n_\text{signal times}$}{
    	    $\mathcal{Y}_\text{ps} \leftarrow$ compute signal time series for pixel $Y$ that we expect if there is a planet at $Y$ at time $t$\;
    		$\mathcal{R} \leftarrow$ get residual time series for pixel $Y$ that was obtained using $t$\;
    		$\text{CS} \leftarrow$ compute cosine similarity of $\mathcal{Y}_\text{ps}$ and $\mathcal{R}$\;
    		\If{ $\text{CS} > \text{maximum}$ }{
    		    $t^* \leftarrow t$\;
    		    maximum $\leftarrow$ CS\;
    		}
    	}
    }
	\caption{Computing the hypothesis time $t^*$ for $Y$} 
	\label{alg:hypothesis-time}
\end{algorithm}

In the next step, we loop over the hypothesis map to compute the \emph{match fraction map}, as outlined in \cref{alg:match-fraction}.
We call this quantity the \emph{match fraction} because it measures the fraction of affected pixels that match the original hypothesis for the target pixel.
We expect that pixels on the planet trajectory receive high match fraction values, while pixels that only contain noise should have low match fractions (see \cref{fig:match-fraction-map}).

\begin{algorithm}[t]
	\SetArgSty{}
	\DontPrintSemicolon
	\setlength{\parskip}{2.3pt}
	{
        $t^* \leftarrow$ get hypothesis time for $Y$ from hypothesis map \;
        $A \leftarrow$ find set of affected pixels (i.e., pixels on the planet path if there is a planet at $Y$ at time $t^*$) \;
        $V$ $\leftarrow$ empty list  \;
        \ForEach{pixel $a$ in $A$}{
            $\mathcal{Y}_\text{ps}^a \leftarrow$ compute signal time series for pixel $a$ that we expect if there is a planet at $Y$ at time $t^*$  \;
            $t \leftarrow$ find time on temporal grid closest to peak of $\mathcal{Y}_\text{ps}^a$  \;
            $\mathcal{R} \leftarrow$ find residual time series for $t$  \;
            CS $\leftarrow$ compute cosine similarity of $\mathcal{R}$ and $\mathcal{Y}_\text{ps}^a$  \;
            Add CS to $V$ \;
        }
        $\text{match fraction} \leftarrow \text{median}(V)$ \;
	}
	\caption{Computing the match fraction for pixel $Y$} 
	\label{alg:match-fraction}
\end{algorithm}

Once we have assembled the full match fraction map, we compute from it a \emph{residual selection mask}.
To this end, we project the match fraction from the standard Cartesian coordinate system (given by the right ascension and declination) to polar coordinates defined by the separation from the image center and the azimuthal angle (see \cref{fig:polar-match-fraction-map}).
In this coordinate system, the signature pattern that we expect to see from a planet (in Cartesian coordinates: an arc with an opening angle matching the field rotation) becomes translation invariant, and we can search for it by cross-correlating the polar match fraction map with the expected template (see \cref{fig:cross-correlation-and-template}).
We then use a standard blob detection algorithm (Laplacian of Gaussian) to find the peaks in the resulting cross-correlation map, and convert the result into the desired residual selection mask (see \cref{fig:residual-selection-mask}).%
\footnote{
    Note: We have described here a fully automatic way to determine the residual selection mask from the match fraction map.
    In practice, however, this step will likely benefit from additional manual supervision to assess the plausibility of the results.
}

Finally, we assemble the residual stack: 
For a pixel \Ypos selected by the residual selection mask, we use the residual time series $\mathcal{R}(Y, t^*)$ (with $t^*$ according to the hypothesis map); otherwise, we use the residual time series from the vanilla HSR.
Once we have assembled the full residual stack, we derotate each frame by its respective parallactic angle and then average along the temporal axis to compute our signal estimate.

\subsection{Hypothesis-based HSR}
\label{subsec:hypothesis-based-hsr}

The previous section has outlined how we can use a modified half-sibling regression approach to perform a blind search, that is, to post-process data sets for which we do not know if they contain a planet.
However, there may also be cases where we already have a strong hypothesis for the position of a (potential) planet, for example, because another post-processing algorithm has produced a detection.
In this case, we can substantially simplify our method and drop the loop over the temporal grid: 
If we have a candidate, we can compute for each position the exact time at which we expect the planet to cross that position, giving us a \enquote{perfect hypothesis map}.
Now we compute only \emph{one} residual time series per pixel, either using vanilla HSR or a signal fitting~/~masking (based on our perfect hypothesis map).

Of course, when the initial hypothesis is wrong, this variant of the HSR method can produce false positives.
We, therefore, suggest using this version with caution and only apply it to estimate the photometry for well-established planets (see \cref{subsec:photometry-real-planets}).

\begin{figure*}
    \centering
    \begin{minipage}[t]{184mm}
    \begin{subfigure}[t]{40mm}
        \centering
        \includegraphics[scale=1]{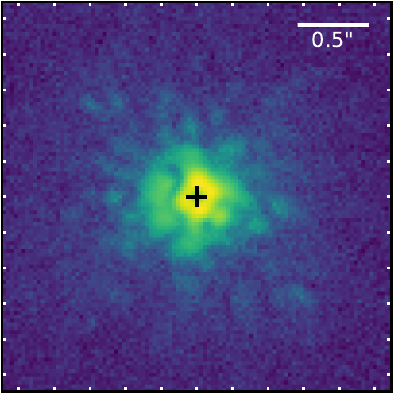}
        \subcaption{Beta Pictoris $L'$}
        \label{fig:beta-pictoris-lp-raw}
    \end{subfigure}%
    \hfill%
    \begin{subfigure}[t]{40mm}
        \centering
        \includegraphics[scale=1]{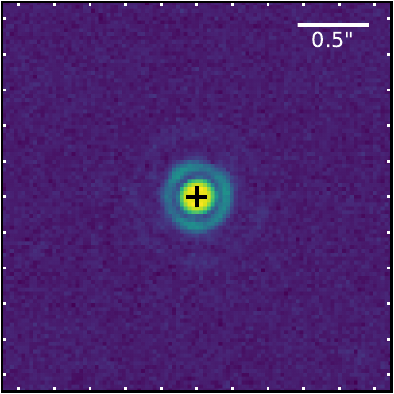}
        \subcaption{Beta Pictoris $M'$}
        \label{fig:beta-pictoris-mp-raw}
    \end{subfigure}%
    \hfill%
    \begin{subfigure}[t]{40mm}
        \centering
        \includegraphics[scale=1]{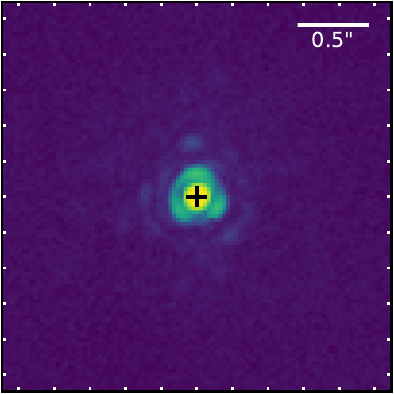}
        \subcaption{HR 8799 $L'$}
        \label{fig:hr-8799-lp-raw}
    \end{subfigure}%
    \hfill%
    \begin{subfigure}[t]{40mm}
        \centering
        \includegraphics[scale=1]{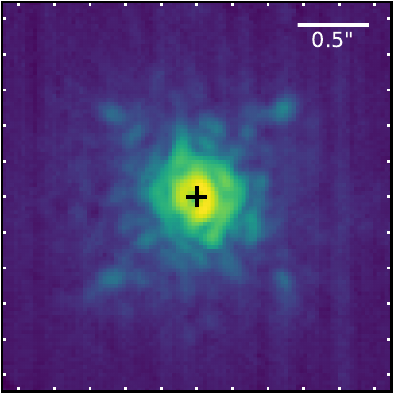}
        \subcaption{R Coronae Australis $L'$}
        \label{fig:r-cra-lp-raw}
    \end{subfigure}%
    \end{minipage}
    \caption{
        Examples of single temporal frames for each of our four data sets.
        To make features at different scales visible, the color bar uses a logarithmic scale.
        For the two data sets that did not use a coronagraph, the frame is dominated by the star at the center.~%
        \href{https://github.com/timothygebhard/hsr4hci/blob/master/scripts/figures/raw-frames/make_plot.py}{\LinkToCode}
    }
    \label{fig:raw-frames}
\end{figure*}

\begin{table*}
\setlength\extrarowheight{2pt}
\setlength{\tabcolsep}{12pt} 
\centering
\caption{
    Details of the four data sets that we use throughout this paper.
}
\label{tab:datasets}
\begin{adjustbox}{max width=\textwidth}
\begin{threeparttable}
\begin{tabular}{lcccc}
    \toprule
    Dataset & 
        Beta Pictoris $L'$ &
        Beta Pictoris $M'$ &
        HR 8799 $L'$ &
        R CrA $L'$ \\ 
    \midrule
    Target star & 
        \simbad{Beta+Pictoris} \object{Beta Pictoris} & 
        \simbad{Beta+Pictoris} \object{Beta Pictoris} & 
        \simbad{HR+8799} \object{HR 8799} & 
        \simbad{R+Coronae+Australis} \object{R Coronae Australis} \\
    Observation date & 
        2013-02-01 & 
        2012-11-26 & 
        2011-09-01 & 
        2018-06-07 \\
    ESO Program ID & 
        \progid{60.A-9800(J)} &
        \progid{090.C-0653(D)} &
        \progid{087.C-0450(B)} &
        \progid{1101.C-0092(A)} \\
    Original reference & 
        \citet{Absil_2013} &
        \citet{Bonnefoy_2013} & 
        ---\tnote{0} & 
        \citet{Cugno_2019} \\
    \midrule
    Instrument & 
        VLT/NACO &
        VLT/NACO &
        VLT/NACO &
        VLT/NACO \\
    Filter / central wavelength & 
        $L'$ (\SI{3.80}{\micro\meter}) &
        $M'$ (\SI{4.78}{\micro\meter}) &
        $L'$ (\SI{3.80}{\micro\meter}) &
        $L'$ (\SI{3.80}{\micro\meter}) \\
    Pixel scale (arcsec / pixel) & 
        0.0271 & 
        0.0271 & 
        0.0271 & 
        0.0271 \\
    $\lambda / D$ (arcsec)\tnote{1} &
        0.0956 &
        0.1202 &
        0.0956 &
        0.0956 \\
    Coronagraph &
        AGPM &
        --- &
        --- &
        AGPM \\
    DIT [science / PSF] (s) & 
        0.2 / 0.020197 &
        0.065 / 0.1\tnote{2} &
        0.2 / 0.2\tnote{2} &
        0.1082 / 0.004256 \\
    FWHM of PSF (pixel) & 
        4.24 & 
        4.97 & 
        4.10 & 
        4.19 \\
    Stack size\tnote{3} &
        $\num{29681} \times 185 \times 185$ &
        $\num{52126} \times 111 \times 111$ &
        $\num{21043} \times 167 \times 167$ &
        $\num{16751} \times 223 \times 223$ \\
    Field rotation ($^\circ$) &
        83.3 &
        51.8 &
        32.5 & 
        121.2 \\
    \midrule
    Separation (mas)\tnote{4} & 
        b: $449.74^{+1.30}_{-1.34}$ &
        b: $458.42^{+3.53}_{-3.51}$ &
        ---\tnote{5} &
        b: $186.7  \pm 1.1$     \\
    Position angle ($^\circ$)\tnote{4} &
        b: $210.26 \pm 0.11$ & 
        b: $211.39^{+0.23}_{-0.24}$ &
        ---\tnote{5} & b: $132.0 \pm 0.2$ \\
    Contrast (mag)\tnote{4} &
        b:   $7.85 \pm 0.06$ &
        b:   $7.64 \pm 0.12$ &
        ---\tnote{5} &
        b:   $6.48 \pm 0.01$ \\
    \bottomrule
\end{tabular}
\begin{tablenotes}
    \footnotesize
    \item[0] Data set previously unpublished.
    \item[1] Computed using the central wavelength for $\lambda$ and $D = \SI{8.2}{\meter}$ for the diameter of the primary mirror.
    \item[2] PSF frames used a neutral density filter (\texttt{ND\_Long}) with a transmission of \SI{2.21 \pm 0.07}{\percent} in the $L'$ and \SI{2.33 \pm 0.10}{\percent} in the $M'$ band; see Table A.2 in \citet{Bonnefoy_2013}.
    \item[3] Format: number of frames (after frame selection) $\times$ frame width in pixels $\times$ frame height in pixels.
    \item[4] Values obtained using MCMC in combination with PCA-based PSF subtraction.
        The values for Beta Pictoris $L'/M'$ are taken from \citet{Stolker_2019}; the values for R CrA $L'$ are taken from \citet{Cugno_2019}.
    \item[5] No published values available for this data set.
\end{tablenotes}
\end{threeparttable}
\end{adjustbox}
\end{table*}

\section{Data sets}
\label{sec:datasets}

To study the properties and performance of our proposed algorithm, we apply it to four publicly available ADI data sets from the Very Large Telescope (VLT) that are known to contain exoplanets.
All four data sets were obtained with the NACO instrument \citep{Lenzen_2003,Rousset_2003}.
We give a more detailed overview in \cref{tab:datasets} and show examples of a single frame in \cref{fig:raw-frames}.

We have chosen to focus our analysis on the $L'$ ($\lambda_\text{central} = \SI{3.80}{\micro\meter}$) and $M'$ ($\lambda_\text{central} = \SI{4.78}{\micro\meter}$) wavelength bands not only because hundreds of archival data sets are readily available, but also because and next-generation HCI instruments for the VLT (ERIS; see \citealt{Davies_2018}) and the ELT (METIS; see \citealt{Brandl_2016}) will be operating in this regime.
Additionally, observations in the $L'$ and $M'$ band usually have detector integration times well below 1 second, which allows us to probe our method also in the presence of short-lived speckles.

Each data set has been prepared using a standard pre-processing pipeline (to perform, \eg dark, flat, and sky subtractions, bad pixel corrections, and frame selection) built with PynPoint \citep{Stolker_2019}.
For the exact details for Beta Pictoris $L'/M'$, see \citet{Stolker_2019}; for R CrA $L'$, see \citet{Cugno_2019}.
The HR~8799~$L'$ data set was obtained through private communications with the first author of \citet{Stolker_2019} who prepared it similarly to the other data sets.
For the experiments on the potential impact of the observing conditions on the post-processing performance, we augment our data sets with interpolated time series of a set of ambient parameters (see \cref{subsec:interpolating-the-oc}).
To foster transparency and improve the reproducibility of our results, we are making our final data sets publicly available.%
\footnote{
    \url{https://doi.org/10.17617/3.LACYPN}
}

\section{Experiments and results}
\label{sec:experiments}

In this section, we describe and discuss a series of experiments in which we study different properties of our proposed algorithm.
For an explanation of the reported performance metrics (\eg the \logfpf score) and how we compute them, see \cref{sec:performance-metrics}.

\subsection{First results and comparison with PCA}
\label{subsec:first-results}

\paragraph{Setup:}
In this first set of experiments, we apply the proposed two-stage version of our HSR-based algorithm to the four data sets from \cref{tab:datasets} and compare the results with those obtained using PCA-based PSF subtraction \citep{Soummer_2012,Amara_2012}.
For all data sets, we choose a circular region of interest (ROI) that is approximately \SI{0.2}{\arcsec} larger than the separation of the outermost planet in the data set.
No temporal binning is applied to the data in this experiment.

We run both the \enquote{signal fitting} and the \enquote{signal masking} version of our method, in both cases using a 3-fold splitting scheme for training and applying the models and with a temporal grid size of 32.
This means that all in all, we learn $3 \cdot (32 + 1)$ models for each pixel in the ROI (the $+1$ is for the vanilla HSR model).
Each such model is trained on approximately 66\% of the available frames and then applied to the hold-out frames to get a prediction.
For the models, we use ridge regression in combination with a leave-one-out cross-validation scheme to determine the value of the regularization parameter, as provided by \href{https://scikit-learn.org/stable/modules/generated/sklearn.linear_model.RidgeCV.html}{\texttt{sklearn.linear\_model.RidgeCV}}. 
We set the possible value range of the regularization strength to the interval $[0.1, 1000]$.

For the PCA-based PSF subtraction, we use the number of principal components suggested in \citet{Stolker_2019} and \citet{Cugno_2019}, that is, 20 components for Beta Pictoris~$L'$ and $M'$, and 9 components for R~Coronae Australis~$L'$.
For the previously unpublished HR~8799~$L'$ data set, we use 20 components.

\begin{figure*}[tp]
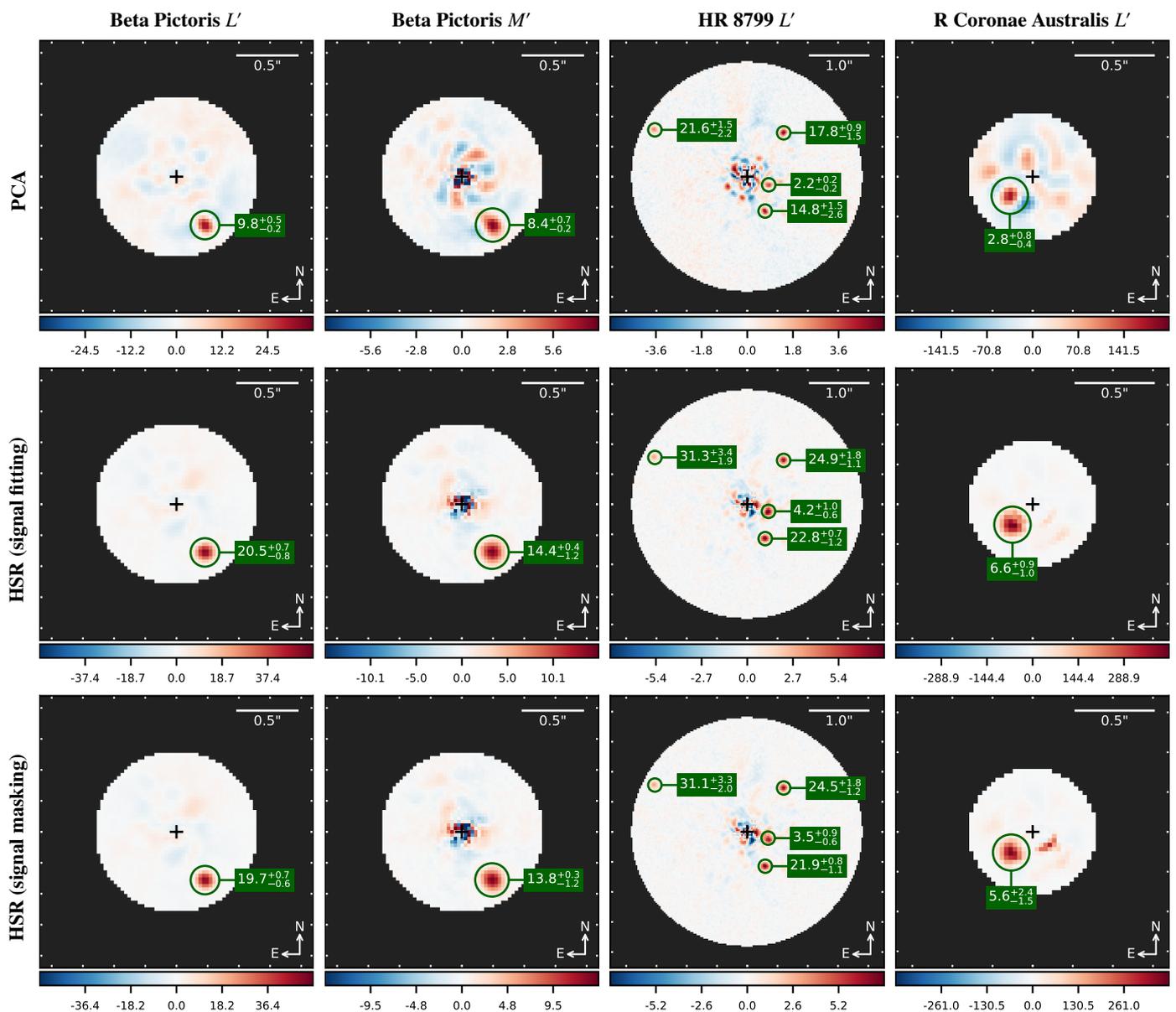

    \centering
    \setlength{\tabcolsep}{0mm}
    \newcommand{\rot}[1]{\adjustbox{margin=1mm}{\rotatebox[origin=c]{90}{\textbf{\small#1}}}}
    \newcommand{\bth}[1]{\adjustbox{margin=1mm}{\textbf{\small#1}}}
    \newcommand{\imginclude}[1]{\adjincludegraphics[valign=M, width=4.3cm, margin=0.9mm]{#1}}
    \begin{tabular}{ccccc}
        & \bth{Beta Pictoris $L'$} & \bth{Beta Pictoris $M'$} & \bth{HR 8799 $L'$} & \bth{R Coronae Australis $L'$} \\
        \rot{PCA} & 
        \imginclude{figures/section-5/5.1_first-results/pca/beta_pictoris__lp} &%
        \imginclude{figures/section-5/5.1_first-results/pca/beta_pictoris__mp} &%
        \imginclude{figures/section-5/5.1_first-results/pca/hr_8799__lp} &%
        \imginclude{figures/section-5/5.1_first-results/pca/r_cra__lp} \\%
        \rot{HSR (signal fitting)} & 
        \imginclude{figures/section-5/5.1_first-results/signal_fitting/beta_pictoris__lp} &%
        \imginclude{figures/section-5/5.1_first-results/signal_fitting/beta_pictoris__mp} &%
        \imginclude{figures/section-5/5.1_first-results/signal_fitting/hr_8799__lp} &%
        \imginclude{figures/section-5/5.1_first-results/signal_fitting/r_cra__lp} \\%
        \rot{HSR (signal masking)} & 
        \imginclude{figures/section-5/5.1_first-results/signal_masking/beta_pictoris__lp} &%
        \imginclude{figures/section-5/5.1_first-results/signal_masking/beta_pictoris__mp} &%
        \imginclude{figures/section-5/5.1_first-results/signal_masking/hr_8799__lp} &%
        \imginclude{figures/section-5/5.1_first-results/signal_masking/r_cra__lp} \\%
    \end{tabular}
    \caption{
        Results of experiment \ref{subsec:first-results}: 
        We compare the output of the PCA-based PSF subtraction approach with two versions of the half-sibling regression-based algorithm on four different data sets obtained with VLT/NACO.
        The images show the signal estimates in units of flux (\ie in the same units like the input data).
        The numbers that label the planets are the respective \logfpf scores.
        The sub- and superscripted values refer to the uncertainties due to the placement of the reference positions used for computing the FPF.
        We find that for all planets, the HSR-based method achieves a higher \logfpf score than PCA.~%
        \href{https://github.com/timothygebhard/hsr4hci/blob/master/scripts/experiments/evaluate-and-plot/evaluate_and_plot_signal_estimate.py}{\LinkToCode}
    }
    \label{fig:first-results}
\end{figure*}

\paragraph{Results:}
When looking at the results shown in \cref{fig:first-results}, we notice several things:
First, for all four data sets and all planets in them, the HSR-based algorithm achieves a higher \logfpf score (implying a lower FPF) than the PCA-based PSF subtraction.
This is, of course, encouraging; however, we need to keep in mind here that the \logfpf score alone does not allow conclusions about the achievable contrast (\ie the detection limits). 

Second, we find that the planet in the HSR-based signal estimates has a rounder shape and lacks the negative \enquote{wings} to the left and right that are common for PCA-based signal estimates.
As these artifacts are typically attributed to self-subtraction, the fact that we do not observe in the HSR-based signal estimates could indicate that our proposed method is less prone to self-subtraction.
We investigate this in more detail in \cref{subsec:photometry-artificial-planets}.

Third, we visually notice that the spatial structure of the residual noise in the signal estimates differs between the PCA- and HSR-based results.
Consequently, if a planet candidate is present in both the PCA and the HSR result, our confidence in a detection may improve. 
We also notice that the two data sets that did not use a coronagraph (Beta Pictoris~$M'$ and HR~8799~$L'$) have the most residual noise close to the center of the image.

Finally, we point out that the absolute scale of the signal estimates values differs between PCA and HSR, with the HSR-based estimates consistently containing a brighter planet.
We will revisit this observation and its potential implications in \cref{subsec:photometry-artificial-planets}.

\subsection{The effect of the choice of predictor pixels}
\label{subsec:choice-of-predictor-pixels}

\paragraph{Setup:}
In \cref{subsubsec:predictor-selection}, we have introduced our choice for the set of pixels that we use as predictors for the model of a given target pixel, which we derived from our domain knowledge discussed in \cref{sec:scientific-domain-knowledge} and \cref{sec:symmetries-in-hci-data-in-practice}.
In this experiment, we now also give an empirical justification for why this choice makes sense.

Our goal, in this case, is essentially to study the properties of the systematic noise and how we can predict it.
Therefore, we begin with taking a data set and removing all known planets from it, which we achieve by injecting an artificial negative planet at the respective positions and contrasts known from the literature.
We do this for all three data sets for which these values are available.
No temporal binning is used.

For each data set, we again choose a region of interest and learn a vanilla half-sibling regression model for each pixel (\ie we do \emph{not} use signal fitting or signal masking) using the usual 3-fold splitting scheme for training and applying the models.
However, instead of using the choice of predictors discussed in \cref{subsubsec:predictor-selection}, we now use \emph{all} pixels that are not part of the exclusion region as predictors.
The base model is again ridge regression, with a regularization strength from the interval $[10, 10^5]$.
We choose a higher regularization strength in this experiment because we have more predictors (\ie more parameters in our model) and because we want to encourage sparser results (\ie many coefficients close to $0$) which allows an easier interpretation.

\begin{figure*}
    \centering
    \begin{minipage}[t]{184mm}
    \begin{subfigure}[t]{60mm}
        \centering
        \hfill%
        \includegraphics[width=29mm]{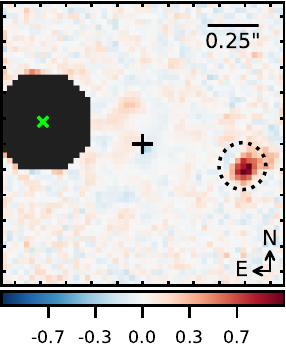}%
        \hfill%
        \includegraphics[width=29mm]{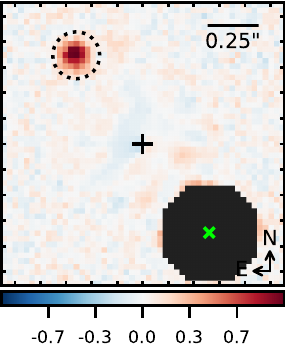}%
        \hfill%
        \subcaption{Examples for Beta Pictoris $L'$.}
    \end{subfigure}%
    \hfill%
    \begin{subfigure}[t]{60mm}
        \centering
        \hfill%
        \includegraphics[width=29mm]{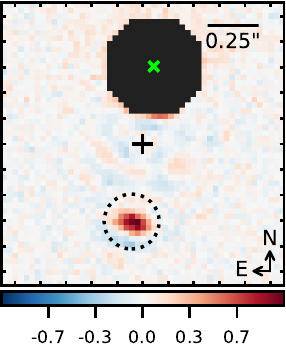}%
        \hfill%
        \includegraphics[width=29mm]{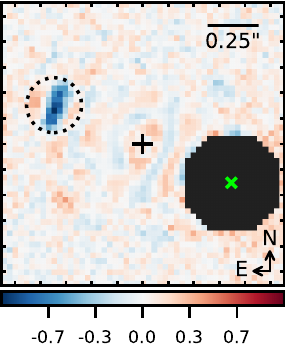}%
        \hfill%
        \subcaption{Examples for Beta Pictoris $M'$.}
    \end{subfigure}%
    \hfill%
    \begin{subfigure}[t]{60mm}
        \centering
        \hfill%
        \includegraphics[width=29mm]{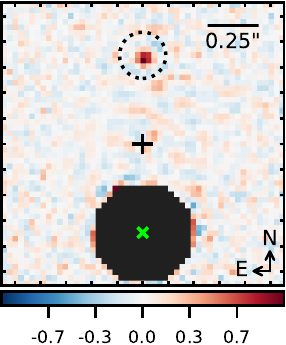}%
        \hfill%
        \includegraphics[width=29mm]{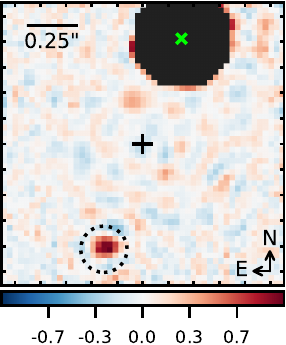}%
        \hfill%
        \subcaption{Examples for R CrA $L'$.}
    \end{subfigure}%
    \end{minipage}
    \caption{
        Example results of experiment \ref{subsec:choice-of-predictor-pixels}.
        Each figure shows a map of the coefficients that constitute the (linear) noise model for the target pixel (green cross).
        The coefficients have been normalized such that the largest absolute value in each frame is 1.
        The dark areas are the respective exclusion regions.
        We notice a familiar pattern: the pixels with the highest (absolute) value---that is, the pixels that contribute most to the prediction of the model---are found in a region that is symmetric across the origin from the target pixel of the model.
        We indicate this region by a dotted circle with a radius equal to 1 FWHM of the respective PSF template.~%
        \href{https://github.com/timothygebhard/hsr4hci/blob/master/experiments/main/5.2_pixel-coefficients/make_plots.py}{\LinkToCode}
    }
    \label{fig:pixel-coefficients}
\end{figure*}

A linear model such as ridge regression consists of one coefficient for each predictor pixel, plus one constant offset.
Therefore, for a given target pixel, we can visualize the model (bar the constant offset) by color-coding each predictor pixel by the value of the corresponding coefficient.
This is a simple way of studying a model's (spatial) structure and investigating which pixels contribute the most to the model's prediction.
Due to the splitting scheme, we technically have three models per pixel.
For plotting purposes, we take the (pixel-wise) mean of the models.
We show exemplary results of these experiments in \cref{fig:pixel-coefficients}.

\paragraph{Results:}
Looking at the model visualizations, we notice a familiar pattern:
The pixels that contribute the most to the prediction of a model for a given pixel \Ypos (in the sense that the coefficients of these pixels have the highest absolute values) are found in a region symmetrically across the origin from $Y$.
This observation matches our insights from \cref{sec:scientific-domain-knowledge} and \cref{sec:symmetries-in-hci-data-in-practice} where we have discussed why we expect the systematic noise in our data to exhibit some degree of spatial symmetry.
The results of this experiment---that is, that the most predictive pixels for the noise at some position $(x, y)$ are the ones in a region around $(-x, -y)$---appear to confirm this expectation well.
(For full disclosure, we note that we do not find this symmetry pattern for \emph{every} target pixel.
However, the pattern is widespread, and for all data sets, it is easy to find examples such as the ones shown in \cref{fig:pixel-coefficients}.)

\subsection{Photometry on signal estimates: artificial planets}
\label{subsec:photometry-artificial-planets}

We know that basic PCA-based PSF-subtraction is prone to over- and self-subtraction: 
The estimate for the systematic noise often contains planet signal which is subsequently removed from the data, resulting in a biased estimate of the photometry that underestimates the planet's brightness \citep{Pueyo_2016}.%
\footnote{
    We refer to \citet{Pueyo_2016} also for a more detailed explanation of the subtle differences between \emph{over-} and \emph{self-}subtraction.
}
If we look at the results in \cref{fig:first-results} again, we notice that the overall scale of the signal estimates is consistently different between the PCA- and the HSR-based results.
In particular, the values of the pixels at the planet positions in the HSR-based signal estimates are, for all four data sets, clearly higher than in the PCA-based results.
This observation motivates the following experiment in which we study whether the half-sibling regression approach produces signal estimates that allow more accurate photometry than a standard PCA-based baseline.

\paragraph{Setup:}
To get a more systematic understanding of the relationship between the photometry estimate and the parameters of the target planet, we run a series of sub-experiments using data into which we inject artificial planets.
For this, we take the Beta Pictoris~$L'$, Beta Pictoris~$M'$ and R~CrA~$L'$ data sets and remove the known companion from each of it by injecting a negative planet using the respective literature values for the position and contrast.
We leave out the HR~8799~$L'$ data set here as there are no literature values for the brightness of the planets.
We store these planet-free data sets and then proceed to create more data sets by injecting artificial planets at different spatial positions and brightness values:
For the contrast between the planet and the star, we choose 15 values between 5 and 12 magnitudes, while for the spatial positions, we use a grid in polar coordinates consisting of 7 values for the distance from the center (2 to 7 FWHM of the PSF) and 6 azimuthal positions (polar angles \SI{0}{\degree}, \SI{60}{\degree}, \ldots, \SI{300}{\degree}), resulting in a total of 630 new data sets for each original data set.
To make the experiments computationally feasible, we apply temporal binning with a binning factor of 128 to all data sets.

\begin{figure}[tbp]
    \centering
    \begin{subfigure}{\linewidth}
        \centering
        \includegraphics[width=90mm]{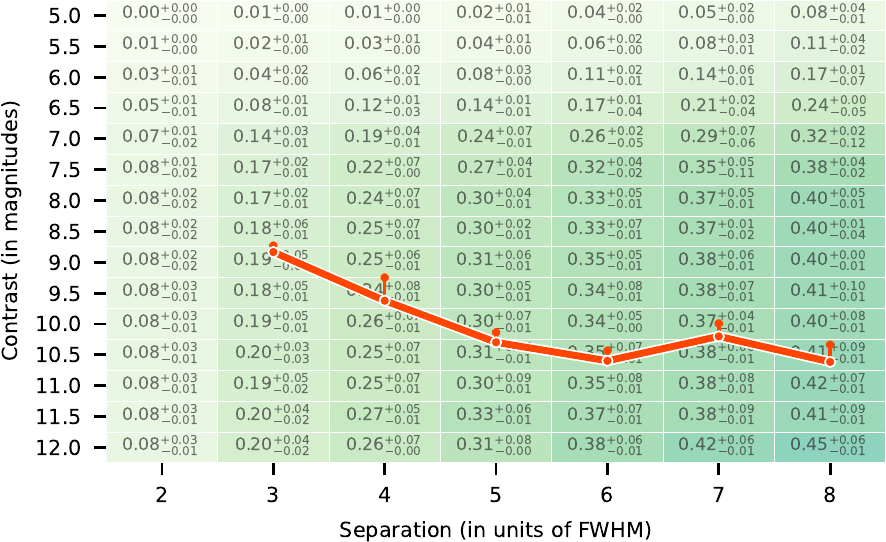}
        \subcaption{Throughput and detection limits for PCA ($n_\text{PC}=20$).}
    \end{subfigure}\\[5mm]
    \begin{subfigure}{\linewidth}
        \centering
        \includegraphics[width=90mm]{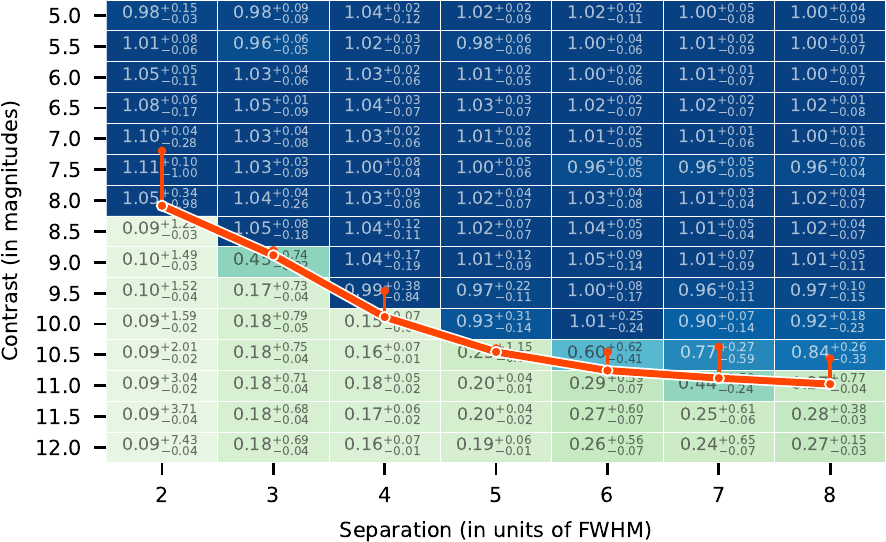}
        \subcaption{Throughput and detection limits for HSR (signal fitting).}
    \end{subfigure}\\[5mm]
    \begin{subfigure}{\linewidth}
        \centering
        \includegraphics[width=90mm]{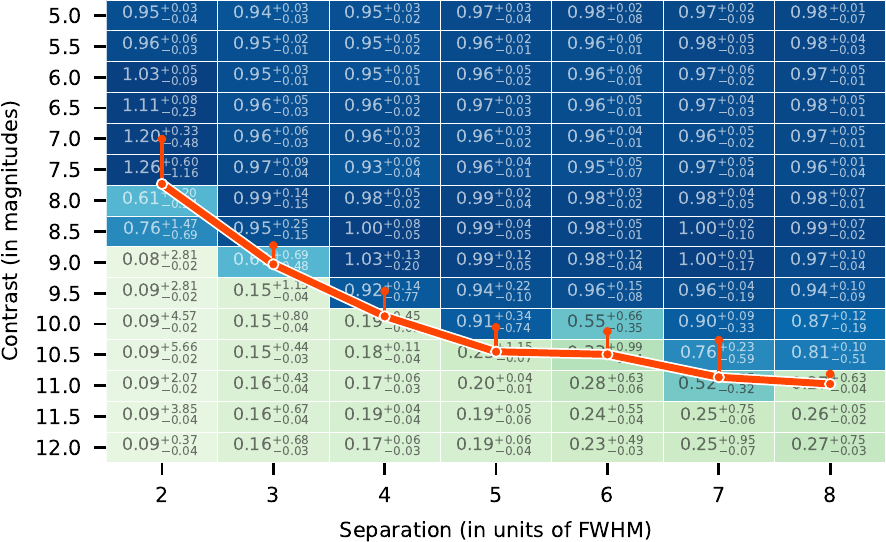}
        \subcaption{Throughput and detection limits for HSR (signal masking).}
    \end{subfigure}\\[5mm]
    \caption{
        Results of experiment \ref{subsec:photometry-artificial-planets} (for the Beta Pictoris $L'$ data set).
        The table plots show the ratio of the observed flux from the signal estimate and the injected flux as a function of contrast and separation, averaged over six azimuthal positions.
        Values close to 1 indicate that the signal estimate provides a good estimate for the planet's brightness.
        Overlaid in orange is the respective $5\sigma$ contrast curve for each method, including a marker that indicates the respective worst-case (\ie the contrast curve that we obtain if we aggregate the data azimuthally by using the position with the highest FPF).
        For PCA, the contrast curve begins at 3 FWHM because, for 2 FWHM, none of the injected planets ever crosses the $5\sigma$ threshold for the FPF.~%
        \href{https://github.com/timothygebhard/hsr4hci/blob/master/experiments/main/5.3_photometry-artificial-planets/02_make_plot.py}{\LinkToCode}
    }
    \label{fig:observed-vs-expected-flux}
\end{figure}

We run both the signal fitting and signal masking versions of our half-sibling regression method on each of the 631 data sets (630 with artificial companions and one without any planet) to produce a signal estimate.
Additionally, we also run all experiments using standard PCA-based PSF subtraction, for three different numbers of principal component: $n \in \lbrace 5, 20, 50 \rbrace$.
For each experiment that contains an artificial planet (\ie every combination of separation, azimuthal position, and contrast), we compute the ratio between the planet flux recovered by the HSR and the true flux value that we used when we injected the artificial planet.
This quantity is sometimes also called \emph{throughput} in the literature.
To this end, we take the signal estimate that we obtain using the injection-free data set and subtract it from the signal estimate that contains an artificial planet.
Then, we measure the flux at the position at which we injected the artificial planet.%
\footnote{
    The motivation for this is that the throughput is commonly defined for denoising models $m$ that are assumed to be linear in the sense that:
    {
    \setlength{\abovedisplayskip}{3pt}
    \setlength{\belowdisplayskip}{3pt}
    \begin{equation*}
        m(\text{noise} + \text{planet signal}) 
        = m(\text{noise})\ +\ \text{throughput} \cdot \text{planet signal} \,,
    \end{equation*}
    }
    where, ideally, $m(\text{noise})$ is close to 0 and the throughput is 1.
    This approach matches, for example, the current implementation of the VIP package (version 1.0.3; \citealt{GomezGonzalez_2017}).
    We note that we have also investigated what happens if we do not subtract the injection-free signal estimate and instead compute an estimate for the residual noise from a set of reference positions at the same separation as the injected companion. 
    We then subtract this estimate for the residual noise from the measured signal before computing the ratio of the observed and expected brightness.
    This approach is more realistic in the sense that it can also be applied to real data sets (\ie not artificially injected planets) where we want to estimate the brightness of a companion.
    We found that the differences between the two approaches were generally very small.
}
Finally, we aggregate the results by averaging azimuthally, giving us a single value for the ratio between the observed and expected planet brightness for each combination of separation and contrast.
We present the final results in \cref{fig:observed-vs-expected-flux}.
For brevity, we are only showing the plots for the Beta Pictoris~$L'$ data set; however, we report that the results for the other two data sets look very similar.

\begin{figure}[tbp]
    \centering
    \includegraphics[width=90mm]{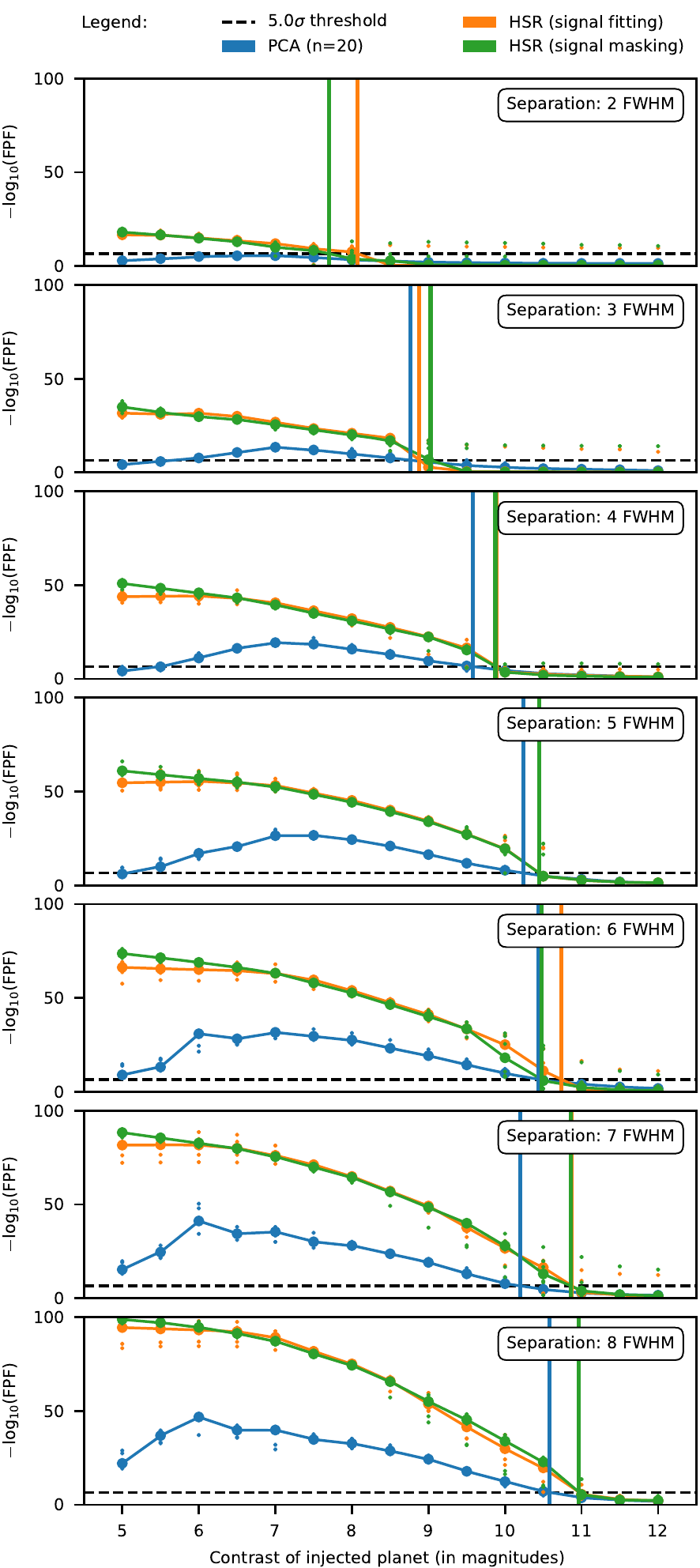}
    \caption{
        To compute the contrast curves shown in \cref{fig:observed-vs-expected-flux}, we linearly interpolate, for each separation, the negative logarithm of the false positive positive fraction (which we compute as the average across 6 azimuthal positions).
        Then, we determine the contrast value at which the \logfpf score crosses the $5\sigma$~threshold and mark these points by a vertical line. 
        This is the contrast curve.
        We note that there is no simple linear relationship between the maximum \logfpf score and the point at which the \logfpf values cross the threshold, demonstrating clearly why the FPF (or SNR) of a single, given planet generally does not allow statements about the achievable detection limits on that data set.~%
        \href{https://github.com/timothygebhard/hsr4hci/blob/master/experiments/main/5.3_photometry-artificial-planets/03_plot_interpolated_fpf.py}{\LinkToCode}    
    }
    \label{fig:fpf-interpolation}
\end{figure}

We also use the output of these experiments to compute detection limits in the form of contrast curves for both PCA and the two HSR versions.
To this end, we first compute the false positive fraction for injected companion and aggregate the results azimuthally.
Then, for each separation, we linearly interpolate the \logfpf as a function of contrast and find the contrast value at which the FPF crosses the threshold of $5\sigma$, where $5\sigma$ refers to the quantiles of a standard normal distribution (\ie we place the threshold at an FPF of approximately 1 in 3.5 million).
We illustrate this procedure in \cref{fig:fpf-interpolation}. 
In some cases (PCA at a separation of 2 FWHM), the interpolated \logfpf never crosses this threshold; in these instances, the respective contrast curve is missing a value.
The final contrast curves are overlaid in \cref{fig:observed-vs-expected-flux}.
Additionally, \cref{fig:all-contrast-curves} shows a direct comparison of all contrast curves, including those that we have computed after repeating the experiments here with the observing conditions as additional predictors for the half-sibling regression (see \cref{sec:observing-conditions}).

At this point, we would like to emphasize that the detection limits that we obtain by this procedure are independent of the throughput values that we have computed in the previous step.
In particular, we do not assume that the throughput is independent of the brightness of the planet signal.

\paragraph{Results:}
Looking at our results, we notice two things.
First, the detection limits obtained by our HSR-based method are, in all cases, comparable or better than those of the PCA-based baseline, with improvements of over one magnitude in the best cases (see also \cref{fig:all-contrast-curves} for easier comparison).

Second, we find that for our method, the ratio of the observed and the expected planet flux is close to 1 for virtually the entire parameter space above the contrast curve. 
This observation suggests that if the HSR is able to detect a planet, it also provides a reasonable estimate for the planet's brightness.
This is in contrast to PCA, where we generally cannot perform meaningful photometry directly on the signal estimate and need additional techniques to correct for the algorithm's bias \citep{Pueyo_2016}.

For HSR, we also notice a sharp drop in the ratio of observed and the expected flux, whose position generally follows the contrast curve.
Closer inspection reveals that this drop corresponds precisely to the point where planets are too faint to produce a detectable signature in the match fraction map.
Consequently, all pixels simply default to their vanilla HSR model, which, as discussed before, suffers from significant over-subtraction.

\subsection{Photometry on signal estimates: real planets}
\label{subsec:photometry-real-planets}

Motivated by the results from the last experiment, we apply the HSR to the three real data sets for which we know the brightness of the planets that they contain (\ie we exclude the HR~8799~$L'$ data set) and compare the contrast estimate based on the HSR signal estimate with the literature values.
For this special case, we do not have to treat the problem as a blind search but can try to use the HSR in \enquote{hypothesis-based mode}, as suggested in \cref{subsec:hypothesis-based-hsr}. 
This means that we use our knowledge about the positions and movement of the planets to only train a single model for each pixel: 
Pixels that are not on the trajectory of the planet are denoised using a vanilla HSR model, whereas for the planets on the planet trace, we use a signal fitting or signal masking-based model.
This significantly reduces the computational cost.

We run the experiment for three amounts of temporal binning, otherwise using the same hyper-parameters as in the previous experiments.
For each signal estimate, we perform photometry at the expected planet position and compare the contrast with the respective literature values.
The results are shown in \cref{tab:contrast-estimates-real-planets}.

\begin{table*}[t]
    \centering
    \tiny
    \caption{
        Results of experiment \ref{subsec:photometry-real-planets}.
        This table shows the estimated contrast in magnitudes when running both versions of our HSR approach in hypothesis-based mode, for three different data sets and temporal binning factors (1, 10, 100). 
        The numbers in parentheses denote the (dimensionless) ratio of the estimated contrast and the literature values in linear units (\ie flux ratios).
        Values in bold fall inside the MCMC-based uncertainty intervals from the literature.~%
        \href{https://github.com/timothygebhard/hsr4hci/blob/master/experiments/main/5.4_photometry-real-planets/get_contrast_estimates.py}{\LinkToCode}
    }
    \label{tab:contrast-estimates-real-planets}
    \begin{minipage}[t]{184mm}
    \begin{tabular*}{\linewidth}{ll@{\extracolsep{\fill}}ccccccccccc}
    \toprule
    & & 
    \multicolumn{3}{c}{Beta Pictoris $L'$} & & 
    \multicolumn{3}{c}{Beta Pictoris $M'$} & & 
    \multicolumn{3}{c}{R CrA $L'$} \\
    \cmidrule(lr){2-5}
    \cmidrule(lr){7-9}
    \cmidrule(lr){11-13}
     &
        &
        1 & 
        10 &
        100 &
        &
        1 &
        10 &
        100 &
        &
        1 &
        10 &
        100 \\
    \midrule
    Signal fitting &
        &
        \textbf{7.79} (1.05) &
        7.76 (1.09) &
        7.73 (1.11) &
        &
        \textbf{7.53} (1.11) &
        \textbf{7.54} (1.10) &
        \textbf{7.53} (1.11) &
        &
        6.85 (0.71) &
        6.79 (0.75) &
        6.82 (0.73) \\
    Signal masking &
        &
        \textbf{7.85} (1.00) &
        \textbf{7.82} (1.03) &
        \textbf{7.81} (1.04) &
        &
        \textbf{7.57} (1.07) &
        \textbf{7.58} (1.06) &
        \textbf{7.58} (1.06) &
        &
        7.01 (0.62) &
        7.24 (0.50) &
        7.33 (0.46) \\
    \bottomrule
    \end{tabular*}
    \end{minipage}
\end{table*}

For both the $L'$ and $M'$-band data sets of Beta Pictoris, we find that the contrast estimates based on the HSR are very close to the values from the literature that were obtained using a combination of PCA and Markov Chain Monte Carlo (MCMC).
Most values even fall within the uncertainty intervals reported in the literature.
It also appears that the temporal binning factor does not strongly impact the estimated contrast values, which is a potentially helpful insight for reducing the computational cost of the method.

For the R~CrA~$L'$ data set, the contrast values based on HSR are not in good agreement the MCMC-based values reported in \citet{Cugno_2019}.
There are, however, several things that we need to keep in mind here:
First, \citet{Cugno_2019} compute their estimates for the astrometry and photometry of R~CrA~$b$ using two different methods, MCMC and Hessian matrix optimization (HM), and results of the two approaches do not agree, neither for the position nor the contrast (see Table 3 in \citealt{Cugno_2019}).
Especially the uncertainty of the position potentially has a significant impact on our result, as it determines the hypothesis map and thus the type of model that we learn for each pixel.
Second, if we compare our contrast values with the HM-based estimate (\SI{6.70 \pm 0.15}{\magnitude}), we find that the HSR-based values are much closer to the literature values. 
For signal fitting, they even fall inside the reported uncertainty interval.
Third, \citet{Cugno_2019} observed the star at two different times in 2017 and 2018, and the contrast values for epoch~1 and epoch~2 do not agree.
(We only use the data from 2018, that is, epoch 2.)
If we also include the contrast values based on epoch~1 in our comparison ($7.29 \pm 0.18$\,mag for HM and $6.93^{+0.01}_{-0.02}$\,mag for MCMC), almost all our contrast values for R~CrA~$b$ match at least one literature value.

\section{Observing conditions as additional predictors}
\label{sec:observing-conditions}

In \cref{subsubsec:predictor-selection}, we argued that the half-sibling regression framework is very flexible and that our choice of predictors for a given target pixel is not limited to other pixels from the same detector but can also include additional quantities that are informative for the systematic noise such as, for example, the observing conditions of a data set. 
Extending the model in this direction builds on the insight that we know that the observing conditions are related to the systematic noise through various effects:
For example, the lifetime of atmospheric speckles depends on the wind speed and seeing inside the control region of the AO system, and the refractive index of the air, which determines the optical path length of the wavefront, depends on the air temperature \citep{Males_2021}.
Furthermore, studies by \citet{Tallis_2018} and \citet{Xuan_2018} have already shown that the observing conditions of a data set are highly predictive of the achievable contrast.
We, therefore, believe that the observing conditions are a natural candidate for showcasing the flexibility of the HSR framework by extending our proposed method to incorporate this additional meta-information into the post-processing procedure.
This section describes how we have prepared the respective data for this and contains the experiments we have performed to test the effect of the observing conditions on the denoising performance in practice.

\subsection{Interpolating the observing conditions}
\label{subsec:interpolating-the-oc}

\begin{figure*}[t]
    \centering
    \includegraphics[width=184mm]{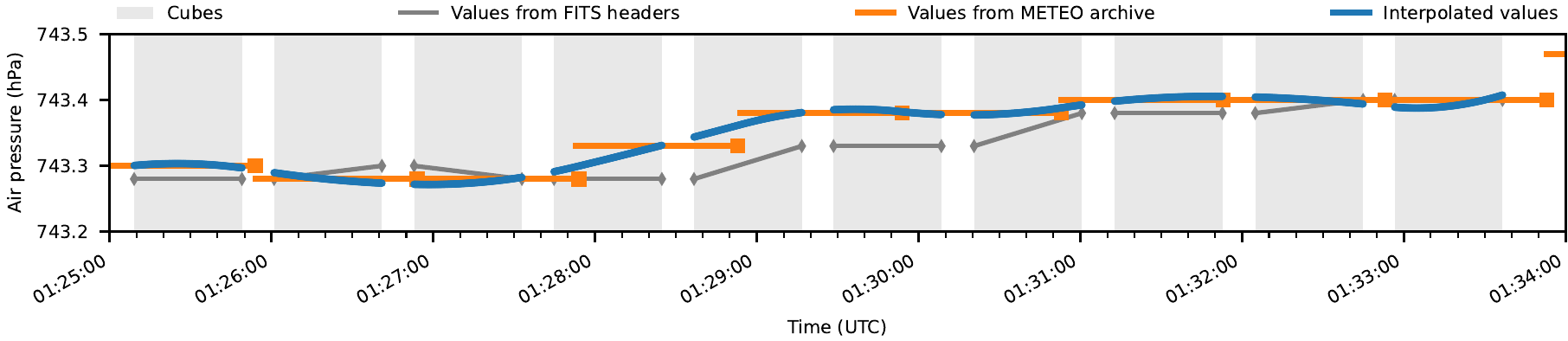}
    \caption{
        Example results of the spline interpolation of the observing conditions (here: air pressure for the Beta Pictoris~$L'$ data set).~%
        \href{https://github.com/timothygebhard/hsr4hci/blob/master/scripts/figures/obscon-interpolation/make_plot.py}{\LinkToCode}
    }
    \label{fig:air-pressure}
\end{figure*}

The VLT routinely records a large number of parameters that quantify the observing conditions and makes them available through a public archive.%
\footnote{\url{https://archive.eso.org/cms/eso-data/ambient-conditions/paranal-ambient-query-forms.html}}
However, these archival observing conditions generally do not have the same temporal resolution as the science data (\ie the images).
Instead, only averages over integration periods that are much longer than the exposure time of a single frame are available:
the observing conditions are typically averaged over intervals of 60 seconds, while detector integration times in the $L'$ and $M'$ band are below 1 second.

If we want to incorporate the observing conditions into the half-sibling regression approach by adding them as additional predictors (see below), we must match their temporal resolution to the science data.
We accomplish this by using a modified spline interpolation approach.
As mentioned above, we do not have access to instantaneous values, which means that we do not know the true value of some parameter $p$ at any time $t$.
Instead, we only know the average of $p$ over intervals $[t_i, t_{i+1}]$:

\begin{align}
    \frac{1}{t_{i+1} - t_i} \int_{t_i}^{t_{i+1}} p(t) \diff t = \bar{p}_i \,.
\end{align}

To upsample $p$, we need to make additional assumptions. 
For example, we might assume that $p(t)$ is continuous and smooth, which seems like a reasonable assumption for most observing conditions.
Under this constraint, we can use $P(t) = \int_{t_0}^{t} p(t') \diff t'$ to rewrite the above equation as:

\begin{align}
    P(t_{i+1}) = P(t_i) + \bar{p}_i \cdot (t_{i+1} - t_i) \,.
\end{align}

If we interpolate this quantity with an appropriate function class (\eg cubic splines) and take the first derivative, we end up with a smooth, continuous time series for $p(t)$ which takes on the desired average values on the original intervals, and which we can evaluate at the observation times of the science frames.
(Of course, the interpolation will still only produce a plausible approximation and not the true value of $p$.)
A practical complication for this last step arises from the fact that some data sets suffer from frame loss, which makes it impossible to determine the time of a given frame precisely and requires us to make further approximations.

In \cref{fig:air-pressure}, we show an example of the result of this interpolation scheme and compare it with the raw archival values as well as the respective values from the FITS files of the data cubes.
The code for querying the ESO archives and interpolating the observing conditions is available as part of our public GitHub repository.

\subsection{Choice of observing conditions}

If we include engineering parameters, more than 100 different ambient condition parameters are available from the ESO archives.
For data sets recorded after April 2016---when the Astronomical Site Monitor (ASM) at Paranal was upgraded---this number is even higher (over 350 parameters).
Of course, many of these parameters are highly correlated or redundant.
Therefore, for this proof-of-principle study, we have decided to use only a relatively small set of parameters inspired by the  headers of the FITS files containing the science data:
\begin{enumerate*}
    \item Air mass$^*$,
    \item Air pressure,
    \item Coherence time $\tau_0$, 
    \item Detector temperature$^*$,
    \item Isoplanatic angle $\theta$,
    \item Primary mirror temperature$^*$,
    \item Observatory temperature, 
    \item Relative humidity,
    \item Seeing, 
    \item Wind speed components $U$, $V$ and $W$.
\end{enumerate*}
Parameters marked with $^*$ are telescope-specific and thus not available from the online archive but only from the headers of the original FITS files, which do not provide a regular temporal grid.
Therefore, we do not apply spline interpolation for these parameters but only interpolate them linearly for each cube.

We suggest that future research should study in greater detail which observing conditions are the most helpful for denoising purposes or which additional pre-processing steps could be applied.
For example, instead of working with the raw observing conditions, one could apply PCA to them and then use the first $n$ principal components as the predictors to minimize redundancy.

\subsection{Incorporating the observing conditions into the HSR}
\label{subsec:incorporating-oc-into-hsr}

For this first proof-of-concept study, we have chosen a conceptually simple and straightforward way to include the observing conditions into the half-sibling regression framework:
Since the observing conditions are time series that have the same form as the time series of a pixel on our detector, we can put the observing conditions on an equal footing with the other predictors and simply add the time series as additional columns to the data matrix \Xts.

\begin{figure}
    \centering
    \includegraphics[width=90mm]{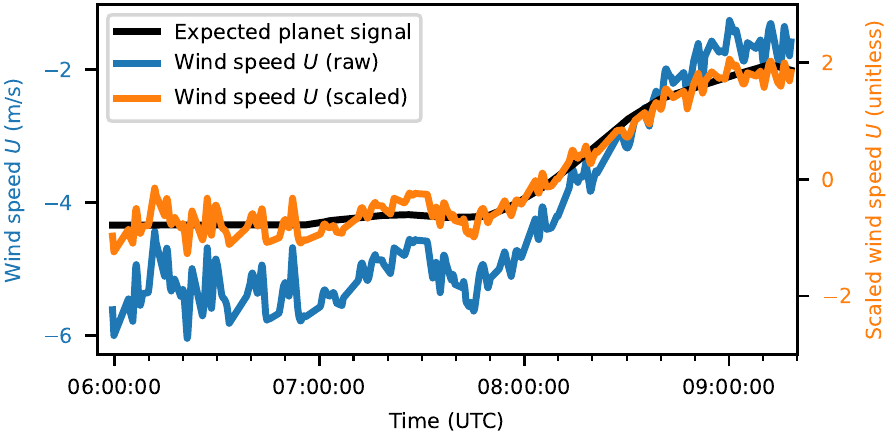}
    \caption{
        This plot shows an example (based on the R~CrA~$L'$ data set) of a significant yet coincidental correlation between an observing condition (in this case: the $U$ component of the wind speed) and the expected signal time series for a pixel that contains a planet at the end of the observation. 
        The correlation coefficient here is \num{0.98}.
        To make the correlation more easily visible, we add a normalized and scaled version of the wind speed (in orange).~%
        \href{https://github.com/timothygebhard/hsr4hci/blob/master/scripts/figures/obscon-vs-expected-signal/make_plot.py}{\LinkToCode}
    }
    \label{fig:obscon-vs-expected-signal}
\end{figure}

There is, however, one caveat: During preliminary experiments, we have found that sometimes, the time series of an observing condition can---by pure chance---mimic the time series of a planet signal, with correlation coefficients reaching absolute values above 0.95.
We show one illustrative example of this in \cref{fig:obscon-vs-expected-signal}.
In these cases, we observed that including the observing conditions as additional predictors can lead to artifacts in the final results and thus deteriorate the overall performance.
As a simple way of preventing this failure mode, we decided to add an extra step to the construction of the data matrix \Xts where we compute the correlation between the observing conditions and the signal time series that we use for signal fitting or signal masking. 
Only if the absolute value of the correlation coefficient is less or equal than $0.3$ do we add the observing condition as an additional predictor.
The value of this threshold was chosen based on a few preliminary experiments but was not systematically optimized.

For our vanilla HSR models (\ie the ones that do not use signal fitting or signal masking), we do not perform this thresholding; instead, we always add all available observing conditions as additional predictors.

\subsection{Experimental setup and results}
\label{subsec:oc-experiment}

\paragraph{Setup:}
To study the effect of the observing conditions, we first repeat the experiments from \cref{subsec:first-results} using a set of predictors that we augment by the observing conditions as described in the previous section.
All other experiment parameters are kept the same.
We show the results of these experiments (in the form of signal estimates) in \cref{fig:obscon-results}.
Additionally, to simplify the comparison between \cref{fig:first-results} and \cref{fig:obscon-results}, we provide an overview of all \logfpf scores in \cref{tab:logfpf-summary}.

In a second step, we also re-run the experiments from \cref{subsec:photometry-artificial-planets} to compute the detection limits with the observing conditions as additional predictors.
We show the results, in which we compare all contrast curves we have computed, in \cref{fig:all-contrast-curves}.

\begin{table*}[t]
	\centering
	\tiny
	\renewcommand*{\arraystretch}{1.5}
	\newcolumntype{L}[1]{>{\raggedright\hspace{0pt}\arraybackslash}p{#1}}
	\newcolumntype{C}[1]{>{\centering\hspace{0pt}\arraybackslash}p{#1}}
	\newcolumntype{R}[1]{>{\raggedleft\hspace{0pt}\arraybackslash}p{#1}}
	    
	\newcommand{\logfpfvalue}[3]{#1 & \raisebox{0.5mm}{\ensuremath{_{#2}^{#3}}}}
	\setlength{\tabcolsep}{0mm} 
	    
	\caption{
		Summary of the \logfpf scores from \cref{fig:first-results} and \cref{fig:obscon-results} (for easier comparison).~%
        \href{https://github.com/timothygebhard/hsr4hci/blob/master/experiments/main/6.4_observing-conditions/make_latex_table.py}{\LinkToCode}
	}
	\label{tab:logfpf-summary}
	
	\begin{tabular}{
		p{30mm}
		p{4.5mm} R{14mm}
		L{14mm}
		p{4.5mm} R{14mm}
		L{14mm}
		p{4.5mm} R{7.5mm}
		L{7.5mm}
		R{7.5mm}
		L{7.5mm}
		R{7.5mm}
		L{7.5mm}
		R{7.5mm}
		L{7.5mm}
		p{4.5mm} R{10mm}
		L{10mm}
		}
		\toprule
		                                                      &   &   
		\multicolumn{2}{C{28mm}}{\textbf{Beta Pictoris $L'$}} &   &   
		\multicolumn{2}{C{28mm}}{\textbf{Beta Pictoris $M'$}} &   &   
		\multicolumn{8}{C{60mm}}{\textbf{HR 8799 $L'$}}       &   &   
		\multicolumn{2}{C{20mm}}{\textbf{R CrA $L'$}} \\
		\cmidrule(lr){3-4}
		\cmidrule(lr){6-7}
		\cmidrule(lr){9-16}
		\cmidrule(lr){18-19}
		                                                      &   &   
		\multicolumn{2}{C{28mm}}{b}                           &   &   
		\multicolumn{2}{C{28mm}}{b}                           &   &   
		\multicolumn{2}{C{15mm}}{b} &
		\multicolumn{2}{C{15mm}}{c} &
		\multicolumn{2}{C{15mm}}{d} &
		\multicolumn{2}{C{15mm}}{e}                           &   &   
		\multicolumn{2}{C{20mm}}{b} \\
		\midrule
		PCA                                                   &   &   
		\logfpfvalue{ 9.8}{-0.2}{+0.5}                        &   &   
		\logfpfvalue{ 8.4}{-0.2}{+0.7}                        &   &   
		\logfpfvalue{21.6}{-2.2}{+1.5} &
		\logfpfvalue{17.8}{-1.5}{+0.9} &
		\logfpfvalue{14.8}{-2.6}{+1.5} &
		\logfpfvalue{ 2.2}{-0.2}{+0.2}                        &   &   
		\logfpfvalue{ 2.8}{-0.4}{+0.8} \\
		Signal fitting                                        &   &   
		\logfpfvalue{20.5}{-0.8}{+0.7}                        &   &   
		\logfpfvalue{14.4}{-1.2}{+0.4}                        &   &   
		\logfpfvalue{31.3}{-1.9}{+3.4} &
		\logfpfvalue{24.9}{-1.1}{+1.8} &
		\logfpfvalue{22.8}{-1.2}{+0.7} &
		\logfpfvalue{ 4.2}{-0.6}{+1.0}                        &   &   
		\logfpfvalue{ 6.6}{-1.0}{+0.9} \\
		Signal fitting + OC                                   &   &   
		\logfpfvalue{27.6}{-1.9}{+4.0}                        &   &   
		\logfpfvalue{25.1}{-0.9}{+1.2}                        &   &   
		\logfpfvalue{42.2}{-6.1}{+6.3} &
		\logfpfvalue{43.9}{-1.9}{+2.5} &
		\logfpfvalue{33.3}{-2.1}{+5.3} &
		\logfpfvalue{17.0}{-2.7}{+2.4}                        &   &   
		\logfpfvalue{ 7.0}{-1.0}{+1.6} \\
		Signal masking                                        &   &   
		\logfpfvalue{19.7}{-0.6}{+0.7}                        &   &   
		\logfpfvalue{13.8}{-1.2}{+0.3}                        &   &   
		\logfpfvalue{31.1}{-2.0}{+3.3} &
		\logfpfvalue{24.5}{-1.2}{+1.8} &
		\logfpfvalue{21.9}{-1.1}{+0.8} &
		\logfpfvalue{ 3.5}{-0.6}{+0.9}                        &   &   
		\logfpfvalue{ 5.6}{-1.5}{+2.4} \\
		Signal masking + OC                                   &   &   
		\logfpfvalue{27.3}{-1.6}{+3.4}                        &   &   
		\logfpfvalue{24.3}{-0.6}{+1.2}                        &   &   
		\logfpfvalue{43.1}{-5.9}{+6.9} &
		\logfpfvalue{43.2}{-2.5}{+3.0} &
		\logfpfvalue{32.9}{-1.9}{+4.9} &
		\logfpfvalue{14.5}{-1.0}{+1.5}                        &   &   
		\logfpfvalue{ 6.9}{-1.5}{+2.7} \\
		\bottomrule
	\end{tabular}
\end{table*}

\begin{figure*}[tp]
    \centering
    \setlength{\tabcolsep}{0mm}
    \newcommand{\rot}[1]{\adjustbox{margin=1mm}{\rotatebox[origin=c]{90}{\textbf{\small#1}}}}
    \newcommand{\bth}[1]{\adjustbox{margin=1mm}{\textbf{\small#1}}}
    \newcommand{\imginclude}[1]{\adjincludegraphics[valign=M, width=4.3cm, margin=0.9mm]{#1}}
    
    \begin{tabular}{ccccc}
        & \bth{Beta Pictoris $L'$} & \bth{Beta Pictoris $M'$} & \bth{HR 8799 $L'$} & \bth{R CrA $L'$} \\
        \rot{Signal fitting} & 
        \imginclude{figures/section-6/6.4_observing-conditions/signal_fitting/beta_pictoris__lp} &%
        \imginclude{figures/section-6/6.4_observing-conditions/signal_fitting/beta_pictoris__mp} &%
        \imginclude{figures/section-6/6.4_observing-conditions/signal_fitting/hr_8799__lp} &%
        \imginclude{figures/section-6/6.4_observing-conditions/signal_fitting/r_cra__lp} \\%
        \rot{Signal masking} & 
        \imginclude{figures/section-6/6.4_observing-conditions/signal_masking/beta_pictoris__lp} &%
        \imginclude{figures/section-6/6.4_observing-conditions/signal_masking/beta_pictoris__mp} &%
        \imginclude{figures/section-6/6.4_observing-conditions/signal_masking/hr_8799__lp} &%
        \imginclude{figures/section-6/6.4_observing-conditions/signal_masking/r_cra__lp} \\%
    \end{tabular}
    \caption{
        Results of experiment \ref{subsec:oc-experiment}.
        The setup is equivalent to experiment \ref{subsec:first-results}, except that we add the observing conditions as additional predictors (see \cref{fig:first-results}, or \cref{tab:logfpf-summary}, for comparison).
        Again, the images show the signal estimates in units of flux (\ie in the same units like the input data) and the numbers that label the planets are the respective \logfpf scores.~%
        \href{https://github.com/timothygebhard/hsr4hci/blob/master/scripts/experiments/evaluate-and-plot/evaluate_and_plot_signal_estimate.py}{\LinkToCode}
     }
    \label{fig:obscon-results}
\end{figure*}

\begin{figure*}
    \centering
    \begin{subfigure}[t]{58mm}
        \centering
        \includegraphics[width=58mm]{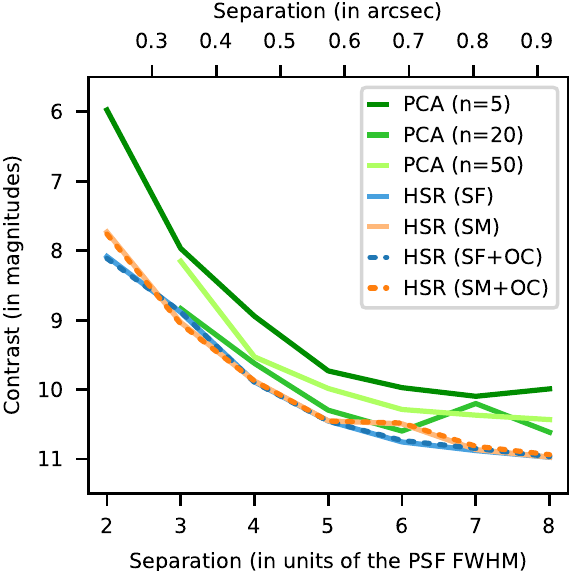}
        \caption{Beta Pictoris $L'$}
    \end{subfigure}%
    \hfill
    \begin{subfigure}[t]{58mm}
        \centering
        \includegraphics[width=58mm]{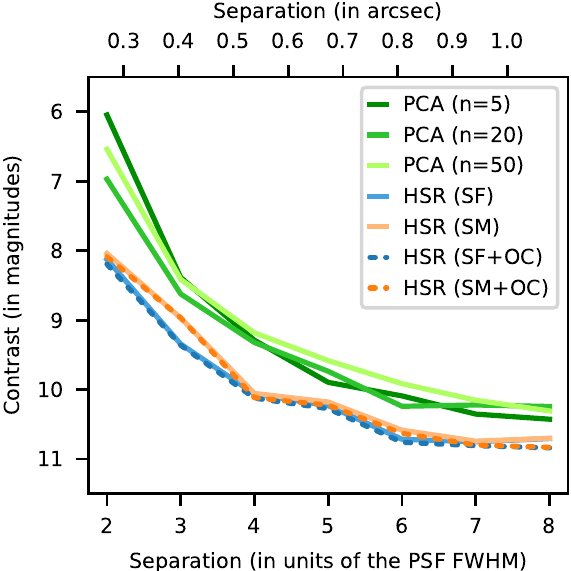}
        \caption{Beta Pictoris $M'$}
    \end{subfigure}%
    \hfill
    \begin{subfigure}[t]{58mm}
        \centering
        \includegraphics[width=58mm]{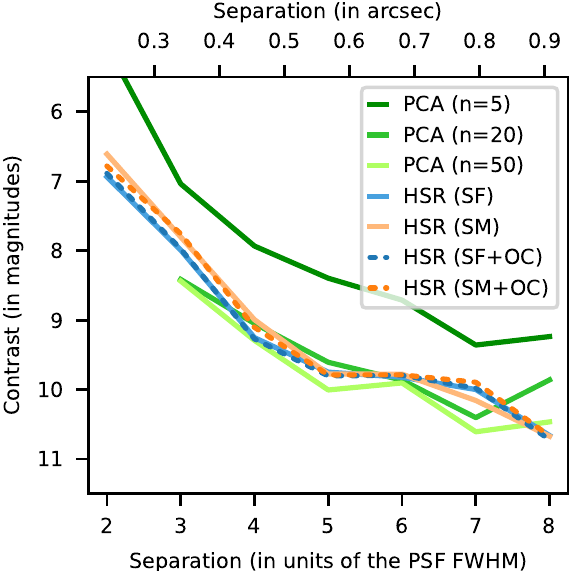}
        \caption{R CrA $L'$}
    \end{subfigure}%
    \caption{
        This figure shows a comparison of all contrast curves that we computed following the procedure described in \cref{subsec:photometry-artificial-planets}.~%
        \href{https://github.com/timothygebhard/hsr4hci/blob/master/experiments/main/5.3_photometry-artificial-planets/04_plot_all_contrast_curves.py}{\LinkToCode}
    }
    \label{fig:all-contrast-curves}
\end{figure*}

\paragraph{Results:}
As a first result, we note that adding the observing conditions as additional predictors in the described way leads to a noticeable and consistent improvement in the \logfpf metric, with factors between approximately \num{1.1} (for the R~CrA~$L'$ data set) and \num{4} (for planet $e$ in the HR~8799~$L'$ data set).

Comparing \cref{fig:obscon-results} with \cref{fig:first-results}, we find that in addition to the \logfpf score improvements, adding the observing conditions as additional predictors also visibly decreases the residual noise at small separations, especially for the data sets that did not use a coronagraph (Beta Pictoris~$M'$ and HR~8799~$L'$).

This visual reduction of the systematic noise does, however, not seem to transfer directly into improved detection limits:
The contrast curve comparison in \cref{fig:all-contrast-curves} suggests that including the observing conditions only has a small impact on the achievable detection limit.
At first, this seems somewhat counter-intuitive.
However, closer investigation reveals that this can likely be explained as an effect of temporal binning: 
While the experiments for \cref{fig:first-results} and \cref{fig:obscon-results} have used unbinned data, the experiments underlying the contrast curves in \cref{fig:all-contrast-curves} are based on data with a temporal binning factor of 128.
However, as our supplementary experiments in \cref{sec:temporal-binning} show, temporally binning the data reduces the usefulness of the observing conditions as additional predictors, and at a binning factor of 128, the signal estimates with and without observing conditions are very similar.
Consequently, it seems consistent that we do not observe a significant difference between the respective contrast curves.

\paragraph{}
We conclude from these experiments that, in general, incorporating the observing conditions as additional predictors for the half-sibling regression can improve our ability to model the systematic noise in HCI data, and we believe that it is a promising direction for future research.
Future work may, for example, study the role of temporal binning in greater detail or look into more sophisticated ways of incorporating the observing conditions than the approach presented here.

\section{Discussion}
\label{sec:discussion}

\subsection{Advantages and limitations of our approach}

In the previous sections, we have seen several advantages of our HSR-based post-processing algorithm for ADI data:
Besides explicitly incorporating prior domain knowledge of the problem, our method is also flexible and easy to extend, as we have demonstrated by adding the observing conditions as additional predictors.
Previous studies (see, \eg \citealt{Ansdell_2018}) have already demonstrated that combining domain knowledge and learning-based systems has great potential for astrophysical applications, and we consider our work an early step in this direction in the field of exoplanet imaging.
Furthermore, our experiments show that our proposed method often produces photometrically well-calibrated signal estimates that allow direct estimation of the contrast between a planet and its host star.
Finally, although the comparison of different post-processing algorithms is explicitly not within the scope of this paper and we want to abstain from making strong claims, we point out that in our experiments, the HSR-based method has produced better signal estimates (in terms of the \logfpf score) and, in almost all cases, better detection limits than the PCA-based baseline.

On the downside, we note that the HSR approach is computationally significantly more expensive than the baseline and has more free (hyper-)parameters that require tuning.
Our method scales approximately linearly with the size of the temporal grid and the number of pixels in the region of interest.
Additional factors that impact the computational cost are the number of frames in the data set, the size of the predictor region, the number of splits for the $k$-fold cross-validation and of course the type of base model.
To get some concrete numbers, we look at a single experiment for the contrast curve estimation (\ie with binned data):
For the Beta Pictoris $L'$ data set, we have 232 frames with a circular region of interest with radius \SI{1.1}{\arcsec} (= 41 pixels).
On a 2019 MacBook Pro 16" featuring a 2.6 GHz 6-Core Intel Core i7 processor, running this experiment with signal fitting takes about 3.5 hours.

Another limitation of our method is that, during the first stage of the method, all pixels are processed independently, even though we know that there exist spatial correlations between neighboring pixels.
(On the other hand, this approach allows for efficient parallelization, thus somewhat alleviating the issue of the computational cost.)
Future work may look into also incorporating these spatial correlations.

\subsection{Signal fitting versus signal masking}

In \cref{subsec:fitting-or-masking}, we introduced two different versions of our method, which we continued to compare throughout the experimental sections.
We find that the difference between the methods was generally small, with the signal fitting variant usually having a slight advantage.
We think this is actually quite remarkable considering that the signal masking method uses strictly less training data than signal fitting.
Furthermore, we note that in the present work, we only used linear base models.
Non-linear models, which are easier to use with signal masking, might improve the results of signal masking beyond the performance of signal fitting.

\subsection{Potential future directions}

Looking ahead, we see various directions for future research based on the results that we have presented in this work.

First, a promising way forward could be to extend the method to work with multi-wavelength data obtained with integral field units (IFUs).
This idea, which is also mentioned in the outlook of \citet{Samland_2021}, should be relatively straightforward to implement and allows to incorporate even more prior domain knowledge, namely the behavior of speckles and planets as a function of wavelength.
The suggestion of \citet{Samland_2021} to learn spatio-temporal models (where the value of a pixel at a time $t$ is also predicted from values of the predictor pixels at times $t' \neq t$) could be another possibility in a similar vein. 
However, we believe that this might be complicated in practice because, for several reasons, the time between two consecutive frames in a data set is often not constant (\eg due to frame selection).

Second, one could explore base models more powerful than (linear) ridge regression to capture more complex, non-linear relationships within the data.
For instance, this could allow us to learn also a potential time variability in the relationships between pixels that we are currently ignoring.
On the downside, using more powerful models would likely further increase the computational cost of the method and make it more susceptible to overfitting.

Third, we believe that we have only just begun to tap into the potential of including the observing conditions into the post-processing of HCI data.
Future research could look into more sophisticated ways of incorporating external information (\eg one global model for the effect of the observing conditions instead of independently adding them into each pixel model) or explore new informative features.
A good candidate for this could be the data of the adaptive optics system, which we know must be confounded with the systematic noise in the data.
In the longer term, this could also have implications for instrument design: The possibility to take into account the observing conditions for post-processing could suggest equipping instruments with additional sensors capturing information about systematic errors.

Finally, a challenging but potentially very rewarding direction could be whether the HSR method lends itself to transfer or continual learning.
Continual learning, in this context, refers to the idea that we might not learn the noise models entirely from scratch for each new data set but instead keep accumulating our knowledge about the data. 
Assuming that the fundamental physics of the processes that generate our data remains the same, for example, for a given instrument and a given filter, each new observation we process would then help us refine and improve our models further. 
This could then be exploited in data sets obtained by large exoplanet imaging surveys comprising up to hundreds of targets.

\section{Summary and conclusion}

In this work, we have presented a new post-processing algorithm for ADI-based exoplanet imaging.
Our method is built with the goal in mind to incorporate the available prior domain knowledge about the data explicitly into the denoising process.
More specifically, we make use of our understanding of the data-generating process as well as the (partially) symmetric structure of the data, for which we have also presented empirical evidence based on real data sets (see \cref{sec:symmetries-in-hci-data-in-practice}).
Experimentally, we find the resulting method to be at least competitive with, if not better than, a PCA-based baseline algorithm that is commonly used by the community. 
Our approach also allows us to include additional metadata in the noise model, for example, the observing conditions, and we showed that this can further improve the results. 
Overall, our results showcase the explicit use of comprehensive scientific knowledge to address the challenges of high-contrast exoplanet imaging and provide first working examples. 
With the number of high-contrast instruments for exoplanet science continuing to increase in the coming years, not only on current \mbox{8--10\,m} telescopes, but also on the future \mbox{30--40\,m} Extremely Large Telescopes, we hope to inspire more research in this direction.

\begin{acknowledgements}

    This research has made use of the services of the ESO Science Archive Facility.
    The authors thank Tomas Stolker and Gabriele Cugno for their help in preparing the data sets.
    The authors also thank the anonymous reviewer whose constructive comments helped to improve this manuscript.
    T.D.G. acknowledges partial funding through the Max Planck ETH Center for Learning Systems. 
    Part of this work has been carried out within the framework of the NCCR PlanetS supported by the Swiss National Science Foundation.

    This research has made use of the following Python packages: 
    \texttt{astropy} \citep{Astropy_2013, Astropy_2018},
    \texttt{astroquery} \citep{Astroquery_2019},
    \texttt{matplotlib} \citep{Matplotlib_2007}, 
    \texttt{numpy} \citep{Numpy_2020},
    \texttt{pandas} \citep{Pandas_2010,Pandas_2020},
    \texttt{photutils} \citep{Photutils_2021},
    \texttt{scikit-image} \citep{ScikitImage_2014},
    \texttt{scikit-learn} \citep{ScikitLearn_2011}, and
    \texttt{scipy} \citep{Scipy_2020}.

\end{acknowledgements}

    \bibliographystyle{aa}
    \bibliography{main.bib}

    \appendix

\section{Evidence for symmetries in ADI data}
\label{sec:symmetries-in-hci-data-in-practice}

As mentioned in \cref{sec:scientific-domain-knowledge}, theoretical studies of the PSF structure suggest that HCI data (and in particular: the speckle pattern) should exhibit some degree of (anti)-symmetry.
One potential caveat here is that the theoretical works on the subject usually derive their statements for particular parameter regimes (\eg in the case of \citet{Perrin_2003}, Strehl ratios above 70\%), which do not necessarily apply to data from current or last-generation instruments and which might call the practical usefulness of the theoretical results into question.
In this short appendix, we perform three simple mini-experiments that provide visual evidence for these symmetry patterns in real data.
The data sets that we use for this are described in detail in \cref{sec:datasets}.

\paragraph{Mini-experiment 1:}
\label{par:mini-experiment-1}

\begin{figure*}
    \centering
    \begin{minipage}[t]{184mm}
    \includegraphics[width=21.25mm]{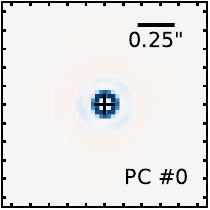}\hfill%
    \includegraphics[width=21.25mm]{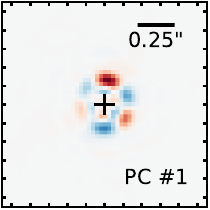}\hfill%
    \includegraphics[width=21.25mm]{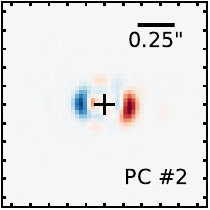}\hfill%
    \includegraphics[width=21.25mm]{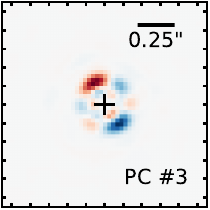}\hfill%
    \includegraphics[width=21.25mm]{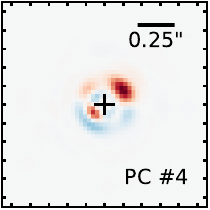}\hfill%
    \includegraphics[width=21.25mm]{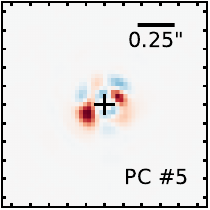}\hfill%
    \includegraphics[width=21.25mm]{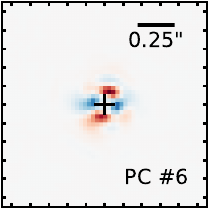}\hfill%
    \includegraphics[width=21.25mm]{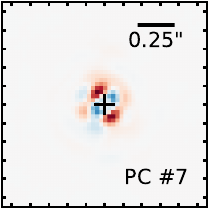}\\[1.5mm]
    \includegraphics[width=21.25mm]{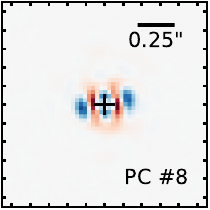}\hfill%
    \includegraphics[width=21.25mm]{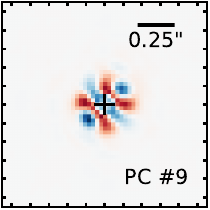}\hfill%
    \includegraphics[width=21.25mm]{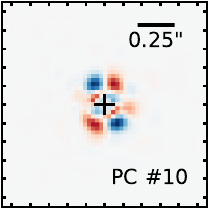}\hfill%
    \includegraphics[width=21.25mm]{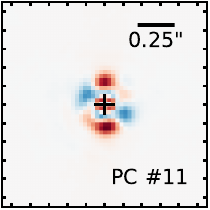}\hfill%
    \includegraphics[width=21.25mm]{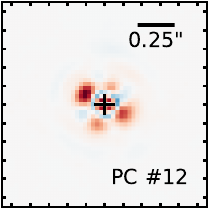}\hfill%
    \includegraphics[width=21.25mm]{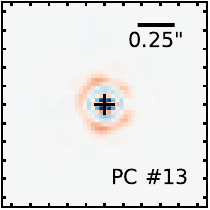}\hfill%
    \includegraphics[width=21.25mm]{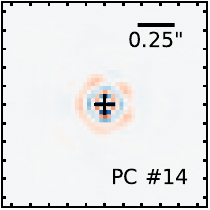}\hfill%
    \includegraphics[width=21.25mm]{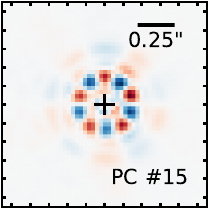}\\[1.5mm]
    \includegraphics[width=21.25mm]{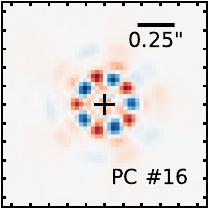}\hfill%
    \includegraphics[width=21.25mm]{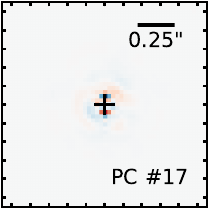}\hfill%
    \includegraphics[width=21.25mm]{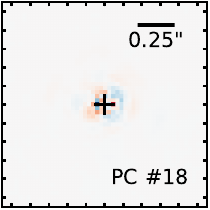}\hfill%
    \includegraphics[width=21.25mm]{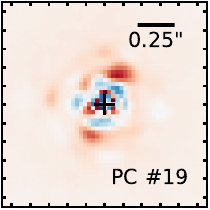}\hfill%
    \includegraphics[width=21.25mm]{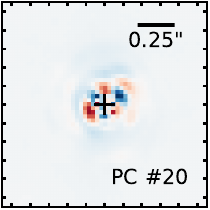}\hfill%
    \includegraphics[width=21.25mm]{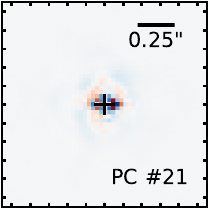}\hfill%
    \includegraphics[width=21.25mm]{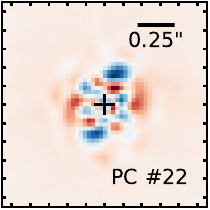}\hfill%
    \includegraphics[width=21.25mm]{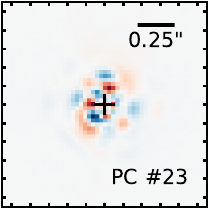}
    \end{minipage}
    \caption{
        Results for mini-experiment 1 (principal components).
        We show the first 24 principal components (ordered by explained variance, in descending order) for the HR 8799 $L'$ data set.
        All images are normalized to the value range $-1$ (blue) to $1$ (red).
        We notice that virtually all principal components exhibit a clear mirror (anti)-symmetry across the origin.~%
        \href{https://github.com/timothygebhard/hsr4hci/blob/master/experiments/appendix/A.1_principal-components/run_experiment_and_make_plots.py}{\LinkToCode}
   }
    \label{fig:principal-components}
\end{figure*}

We perform standard PCA-based PSF-subtraction on a data set and subsequently visualize the principal components that we have computed.
Exemplary results are found in \cref{fig:principal-components}.
We note that many of these eigenimages show a distinct (anti)-symmetry across the origin.
This is interesting because, in PCA-based PSF-subtraction, the estimate for the systematic noise is essentially a weighted sum of these principal components, which means that the noise estimate from PCA will generally also be (imperfectly) (anti)-symmetric.

\begin{figure*}
    \begin{minipage}[t]{184mm}
    \begin{minipage}[t]{90mm}
        \centering
        \begin{subfigure}[t]{43mm}
            \centering
            \includegraphics[width=43mm]{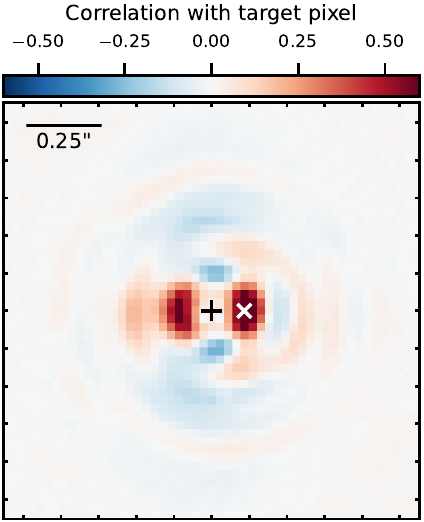}
            \subcaption{Beta Pictoris $M'$}
        \end{subfigure}%
        \hfill%
            \begin{subfigure}[t]{43mm}
            \centering
            \includegraphics[width=43mm]{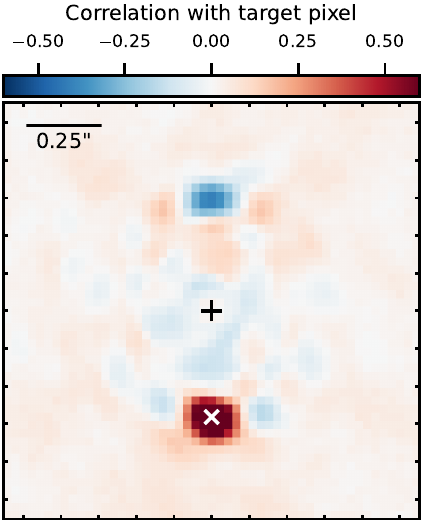}
            \subcaption{HR 8799 $L'$}
        \end{subfigure}%
        \caption{
            Example results for mini-experiment 2 (correlation coefficients).
            We color-code the correlation coefficient of every pixel with a given target pixel (marked by a white cross).
            We see that the target pixels are strongly (anti)-correlated with their immediate neighborhood but also with a similarly sized region symmetrically across the center.~%
            \href{https://github.com/timothygebhard/hsr4hci/blob/master/experiments/appendix/A.2_correlation-coefficient-maps/make_plots.py}{\LinkToCode}
        }
        \label{fig:correlation-maps}
    \end{minipage}%
    \hspace{1mm}%
    \hfill%
    \hspace{1mm}%
    \begin{minipage}[t]{90mm}
        \centering
        \begin{subfigure}[t]{43mm}
            \centering
            \includegraphics[width=43mm]{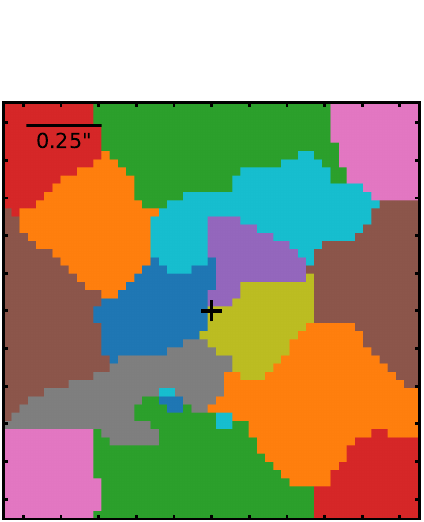}
            \subcaption{Beta Pictoris $L'$}
        \end{subfigure}%
        \hfill%
        \begin{subfigure}[t]{43mm}
            \centering
            \includegraphics[width=43mm]{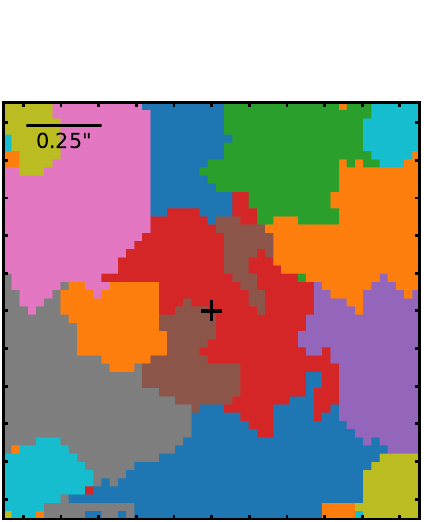}
            \subcaption{R CrA $L'$}
        \end{subfigure}
        \caption{
            Example results for mini-experiment 3 (clustering).
            Clusters are color-coded: pixels with the same color belong to the same cluster.
            (Note: the two examples are independent.)
            Again, we see clear symmetry patterns: pairs of pixels that are symmetric across the origin tend to be part of the same cluster, indicating that the time series of these pixels are similar.~%
            \href{https://github.com/timothygebhard/hsr4hci/blob/master/experiments/appendix/A.3_k-means-clustering/run_experiment_and_make_plots.py}{\LinkToCode}
        }
        \label{fig:k-means}
    \end{minipage}
    \end{minipage}
\end{figure*}

\paragraph{Mini-experiment 2:}
\label{par:mini-experiment-2}
We loop over all spatial pixels $Y = (x, y)$ in a given data set and compute the correlation (along the temporal axis) between $Y$ and every other pixel $Y'$.
We then visualize the results for a given $Y$ by color-coding all pixels by their correlation coefficients.
Unsurprisingly, we find that every pixel $Y$ is strongly correlated with its immediate neighbors.
More interestingly, however, we notice that for many pixels $Y$, there is also a region around the respective mirror-symmetric position $(-x, -y)$ that is clearly (anti-)correlated with $Y$.
We show two examples of such \enquote{correlation maps} in \cref{fig:correlation-maps}.

\paragraph{Mini-experiment 3:}
\label{par:mini-experiment-3}
We take the time series of pixels and treat them as vectors with $T$ dimensions.
We whiten the time series (\ie apply a z-transform to ensure everything is on the same scale) and then apply a simple clustering algorithm (standard $k$-means, as provided by \href{https://scikit-learn.org/stable/modules/generated/sklearn.cluster.KMeans.html}{\texttt{sklearn.cluster.KMeans}}, with $n=10$ clusters) to them, which groups the time series based on their similarity (\ie similar time series end up in the same cluster).
Finally, we color-code each spatial pixel based on the cluster to which its time series was assigned.
Two example results are shown in \cref{fig:k-means}.
Again, we notice a clear pattern: most clusters (indicated by colors in the plot) consist of two regions that are (approximately) symmetric across the origin.

\paragraph{}
We conclude from these mini-experiments that the HCI data sets we are working with do exhibit a noticeable amount of (anti)-symmetry that matches our expectation from theoretical considerations of the PSF structure.     %

\section{Half-sibling regression in detail}
\label{sec:hsr-in-detail}

In this appendix, we explain the idea of half-sibling regression (and its application to ADI data) in more detail and link it, also in terms of notation, to the original work by \citet{Schoelkopf_2016}.

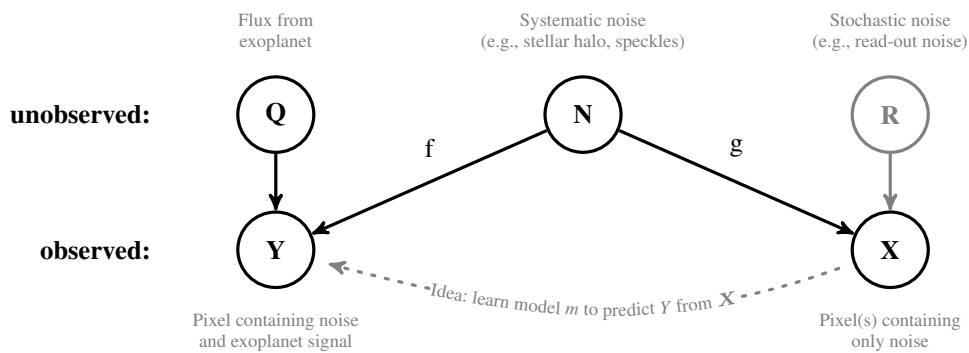
\begin{figure*}
    \centering
    \begin{tikzpicture}[auto, >=stealth']

    \tikzstyle{circ}=[draw=black, very thick, circle, minimum size=1.0cm, inner sep=0, anchor=center, font=\bfseries]
    \tikzstyle{arr}=[->, very thick, draw=black]
    \tikzstyle{annot}=[text=black!50, font=\scriptsize, align=center]

    \draw [fill=white, draw=none] (-8, -3) rectangle (8, 1.5);

    \node [circ, draw=black] at (0, 0) (N) {N};

    \node [circ, left=3cm of N,  draw=black] (Q) {Q};
    \node [circ, below=0.75cm of Q, draw=black] (Y) {Y};
    \node [circ, below=0.75cm of N, draw=none] (M) {};
    \node [circ, right=3cm of N, draw=black!50, text=black!50] (R) {R};
    \node [circ, right=3cm of M, draw=black] (X) {X};
    
    \node [above=2mm of Q, annot] {Flux from\\ exoplanet};
    \node [above=2mm of N, annot] {Systematic noise\\ (e.g., stellar halo, speckles)};
    \node [above=2mm of R, annot, text=black!50] {Stochastic noise\\ (e.g., read-out noise)};
    \node [below=2mm of Y, annot] {Pixel containing noise\\ and exoplanet signal};
    \node [below=2mm of X, annot] {Pixel(s) containing\\ only noise};

    \node [left=1cm of Q, align=right, anchor=east, font=\bfseries] 
        (unobserved) {unobserved:};
    \node [left=1cm of Y, align=right, anchor=east, font=\bfseries] 
        (observed) {observed:};

    \draw [arr] (Q) edge (Y);
    \draw [arr] (N) edge node [midway, above=2mm] {g} (X);
    \draw [arr] (N) edge node [midway, above=2mm] {f} (Y);
    \draw [arr, draw=black!50, fill=black!50] (R) edge (X);
    
    \draw [arr, draw=black!50, loosely dotted, line cap=round, shorten >=0.2cm, shorten <=0.2cm] 
        (X) to [out=200, in=345] (Y);
    \draw[draw=none, fill=white] (-1.95, -2.8) rectangle (2.0, -2.0);
    \draw [decorate, black!50, decoration={text effects along path, text={Idea: learn model {$m$} to predict {$Y$} from {$\bm{X}$}}, text align=center, text effects/.cd, text along path, every character/.style={color=black!50, font=\scriptsize, yshift=-0.55ex}}] (Y) to [out=345, in=200]  (X);

\end{tikzpicture}
    \caption{
        This graph illustrates the causal model that we assume for our data which underlies the idea of half-sibling regression.
    }
    \label{fig:half-sibling-regression}
\end{figure*}

Assume we have a quantity of interest $Q$, which we cannot directly observe.
In our particular case, this would be the photon flux from an extrasolar planet. 
We can, however, observe another \enquote{proxy} quantity $Y$; for example, a pixel on the telescope sensor that contains photons from the planet.
(Formally, $Y$ is a random variable in this setting.)
This pixel $Y$ is also influenced by another latent quantity $N$, which we call the \emph{(latent) systematic noise}. 
We assume the interaction between $Q$ and $N$ to be additive: 
\begin{align}
    Y = Q + f(N) \,, 
    \label{eq:additive-noise-model}
\end{align}
where $f$ is some function that maps the latent systematic noise to the observed systematic noise, that is, $f(N)$ corresponds to the stellar halo, the speckle noise, and generally all instrument effects that we would like to remove from the data. 
For many types of systematic noise, this assumption of an \emph{additive noise model} seems physically justified because the photons of the planet and the noise (\eg speckles or the stellar halo) should indeed simply add up in the detector.
Finally, we consider another pixel $X$ on the sensor that is also affected by $N$ (through a function $g$ which may be different from $f$).
Importantly, both $N$ and $X$ are assumed to be \emph{causally independent} from $Q$: $(X, N) \indep Q$.
(See also \cref{fig:half-sibling-regression} for a graphic illustration of the causal model that we have described here.)
For $N$, this assumption does not require a big leap of faith, as it is hard to imagine how the systematic noise in a telescope would have anything to do with whether or not a given pixel contains a planet.
For $X$, we can make use of the fact that a planet signal on the sensor has a finite spatial size (given by the PSF of the instrument). 
This means that for a given pixel $Y$ that contains $Q$, we can ensure $X \indep Q$ by choosing an $X$ that is sufficiently far away from $Y$.
Note here that for the special case of ADI, one might also want to consider that the planet signal is moving over the sensor during an observation.

The central idea of half-sibling regression is now the following: 
$X$ and $Y$ share some information, but only due to the effect of the unobserved confounder $N$.
However, it is precisely the effect of $N$ on $Y$, that is, $f(N)$, that we would like to remove from $Y$ to find $Q$.
\citet{Schoelkopf_2016} have shown that this is possible in the following way, assuming that the causal model we have just described holds.
If there exists a function $\psi$ such that $\psi(X) = f(N)$ (\ie $X$ contains, in principle, \emph{full information} about the effect $f(N)$), then we can learn a model $m$ that predicts $Y$ from $X$ (\ie we find $\mathbb{E}(Y\,|\,X)$), and we can obtain the following estimate $\hat{Q}$ of $Q$:
\begin{align}
    \hat{Q} = Y - \mathbb{E}(Y\,|\,X) = Y - m(X) = Q - \mathbb{E}(Q) \,.
\end{align}
In other words, by regressing $Y$ onto $X$ and subsequently subtracting the prediction of this regression from $Y$, we can obtain an estimate for $Q$ that is correct up to a constant offset.

Learning a model $m$ that approximates the conditional expectation $\mathbb{E}(Y|X)$ is typically formulated as an optimization problem.
To this end, one assumes a loss function $\mathcal{L}: \mathbb{R} \times \mathbb{R} \to \mathbb{R}$ that measures the distance between the value of $Y$ and the model prediction $m(x)$.
A common choice is the squared difference, $\mathcal{L}(y, x) = (y - m(x))^2$.
We use capital letters here to denote random variables ($Y$, $X$) and lower-case letters for their values ($y$, $x$).
The model $m$ has a set of learnable parameters $\bm{\beta}$, which we find by minimizing the average value of loss function over a set of different realizations of $Y$ and $X$.
In our case, these different realizations simply correspond to the different time steps (or frames), and learning the model $m$ means finding the argument of the minimum:
\begin{equation*}
    \bm{\beta}^* = \argmin_{\bm{\beta}} \sum_{t=1}^{T} \mathcal{L}(y_t, m(x_t ; \bm{\beta}))
\end{equation*}
This, of course, assumes that the relationship between $X$ and $Y$ is constant in time, which is likely not exactly fulfilled in practice.

Furthermore, in practice, the assumption of the existence of a function $\psi$ such that $\psi(X) = f(N)$ is quite strong:
It seems very unlikely that a single pixel will contain all information about the systematic noise so that we can reconstruct $f(N)$.
Luckily, however, we can relax this assumption and still obtain an interesting result.
Let us, therefore, assume that $X$ does \emph{not} contain complete information about $N$. 
However, instead of only observing a single pixel $X$, let us now look at a \emph{set} of $d$ pixels $\bm{X}_d = \{ X_1, X_2, \ldots, X_d \}$ which are all influenced by the same systematic noise $N$ (albeit through possibly different functions $g_{i = 1, \ldots, d}$), but do not contain any contribution from the planet signal $Q$.
Additionally, each $X_i \in \bm{X}_d$ (and also $Y$) may also be affected by stochastic noise terms $R_i$, such as photon shot noise or read-out noise. 
The $R_i$ do not all need to follow the same distribution; however, we assume $R_i$, $N$, $Q$ to be all pairwise disjoint.
Under these assumptions, and a few more rather technical (but reasonable for our application) assumptions about the properties of $f$ and the $g_i$ (in particular, their invertibility) as well as the variance of the $R_i$, \citet{Schoelkopf_2016} show that the following asymptotic result holds:
\begin{align}
    \hat{Q}_d \xrightarrow{\ell_2} Q - \mathbb{E}(Q) \quad \text{for} \quad d \to \infty \,,
\end{align}
where:
\begin{align}
    \hat{Q}_d = Y - \mathbb{E}(Y\,|\,\bm{X}_d)\ \hat{=}\ Y - m(\bm{X}_d) \,.
\end{align}
The intuition for this result is that \enquote{with increasing number of variables, the independent $R_i$ \enquote{average} out, and thus it becomes easier to reconstruct the effect of $N$} \citep{Schoelkopf_2016}.
In practice, of course, the number of possible predictors is always limited, and the speed of the convergence may depend on many factors, including the data set, the model class $m$, and, not least, the particular choice of $\bm{X}_d$.
Whether or not the $\hat{Q}_d$ provides a \enquote{good} estimate of the planet signal will, therefore, depend on the specific science case, and we can only test it experimentally.

\section{Performance metrics for experiments}
\label{sec:performance-metrics}

As noted in \cref{subsec:state-of-the-art}, comparing post-processing algorithms for ADI data in a scientifically meaningful way is hard and existing metrics are limited in various ways.
Nevertheless, some of our experiments require us to quantify the performance of our algorithm.
In this appendix, we explain our respective choice of metrics for these cases and the motivation behind them.
Many of the ideas outlined here will be introduced more thoroughly in an upcoming work by \citet{Bonse_inprep}.

\paragraph{The \logfpf score}
We have chosen to report the \emph{negative decimal logarithm of the false positive fraction} (FPF), denoted as \logfpf, instead of the signal-to-noise ratio (SNR), because the interpretation of the (logarithmic) FPF is independent of the location of the planet candidate:
The SNR, as defined by Eq. (9) of \citet{Mawet_2014}, is based on a set of reference apertures, and the statistical significance depends on the number of these reference apertures.
For example, an SNR of $5$ at a separation of $8\,\lambda/D$ implies a much higher significance than an SNR of $5$ at a separation of $1\,\lambda/D$.
The FPF---given by Eq. (10) of \citet{Mawet_2014}---corrects for the effect of the number of reference positions, thus allowing us to compare values between different planet positions.
The decision to take the negative logarithm of the FPF is motivated by the fact that the \logfpf usually falls into a \enquote{convenient} value range; that is, it typically takes on values between 1 and 50, where \enquote{higher = better}. 
This makes the \logfpf easier to compare than the raw FPF.

\begin{figure}
    \centering
    \includegraphics[width=90mm]{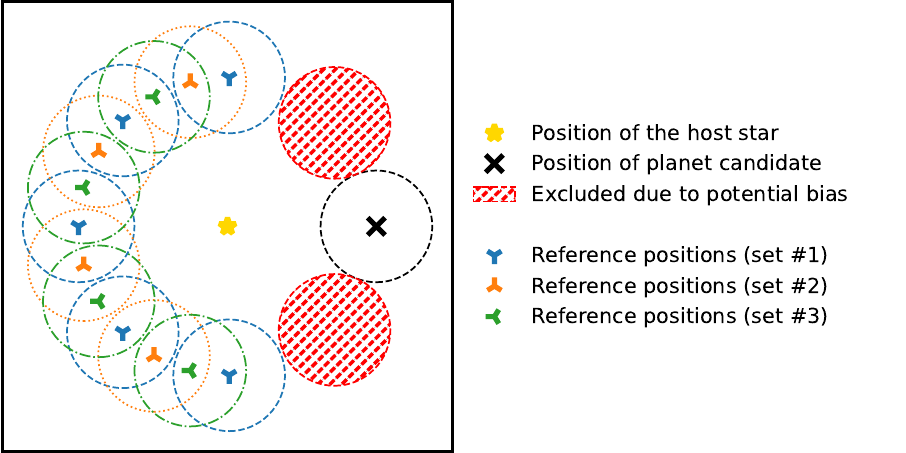}
    \caption{
        To compute the SNR and FPF at a given position, we estimate the noise using $n$ sets of reference positions (here: $n=3$).
        The diameter of the circles is equivalent to 1 FWHM of the PSF template, which is not the size of the apertures used, but was chosen to ensure that the reference positions within each set are independent of each other, in the sense that they are further apart than the typical correlation length (which is given by the size of the PSF).
        We compute the SNR (and FPF) for each set of reference positions before averaging the results across all sets.~%
        \href{https://github.com/timothygebhard/hsr4hci/blob/master/scripts/figures/reference-positions/make_plot.py}{\LinkToCode}
    }
    \label{fig:reference-positions}
\end{figure}

\paragraph{Computation of the FPF}
To compute the SNR (and from it the FPF), we use a procedure that slightly deviates from the approach by \citet{Mawet_2014}.
Most importantly, we do not perform photometry using apertures with a size of $1\,\lambda/D$.
Instead, to measure the \emph{signal} (\ie the flux of the planet), we fit the signal estimate at the position of a planet candidate with a two-dimensional, symmetric Gaussian and integrate the fit result over a disk with a diameter of 1 pixel centered on the mean of the Gaussian.
(The integral has a closed-form solution requiring only the amplitude and the standard deviation of the Gaussian.)
To estimate the noise, we place apertures with a diameter of 1 pixel at the same reference positions that the standard SNR computation uses (\ie at the same separation from the center as the candidate; with azimuthal separations of 1 FWHM of the PSF template).
We omit the positions immediately to the left and the right of the planet candidate because, especially for PCA-based PSF subtraction, they are often affected by self-subtraction effects and thus do not provide an unbiased estimate of the residual noise (see \cref{fig:reference-positions} for an illustration).
Using the values for the signal and the noise that we have obtained as described, we apply Eqs. (9) and (10) from \citet{Mawet_2014} to compute the SNR and the FPF.

We have chosen this \enquote{pixel-based} approach and not the original aperture photometry used by \citet{Mawet_2014}, because we have found that summing over $1\,\lambda/D$-sized apertures can sometimes produce rather non-intuitive results when the residual noise still has spatial structure.
For a more detailed look at this \enquote{pixels versus apertures} discussion, see \citet{Bonse_inprep}.

\paragraph{Placement of reference positions}
We have noticed that the exact placement of the reference positions---which, fundamentally, is arbitrary---can in some cases have a significant impact on the resulting SNR and FPF.
For this reason, we are extending the metric computation in the following way:
Instead of only computing a single estimate for the \enquote{noise} term of the SNR, we compute multiple estimates, each for a slightly different choice of reference positions.
We achieve this by reducing the number of reference positions by one and then \enquote{rotating} them around the image center.
This is illustrated in \cref{fig:reference-positions}.
For each placement of the reference positions, we then compute the SNR and the FPF.
Finally, we take the average of the results across all reference positions.
For more information, see \citet{Bonse_inprep}.

\section{The effect of temporal binning}
\label{sec:temporal-binning}

Compared to PCA-based PSF subtraction, the half-sibling regression approach to post-processing ADI data is computationally more expensive.
A straightforward way to reduce the computational cost is to bin the data temporally by replacing blocks of multiple consecutive frames with their temporal mean.
We can also think of this procedure as increasing the \enquote{effective exposure time} per frame (while decreasing the number of frames).
In this appendix, we present two supplementary experiments that study the effect that this temporal binning has on the performance of our proposed algorithm, in particular in combination with using the observing conditions as additional predictors.

\subsection{The FPF as a function of temporal binning}
\label{subsec:fpf-as-function-of-temporal-binning}

\begin{figure*}[p]
    \centering
    \includegraphics[width=\linewidth]{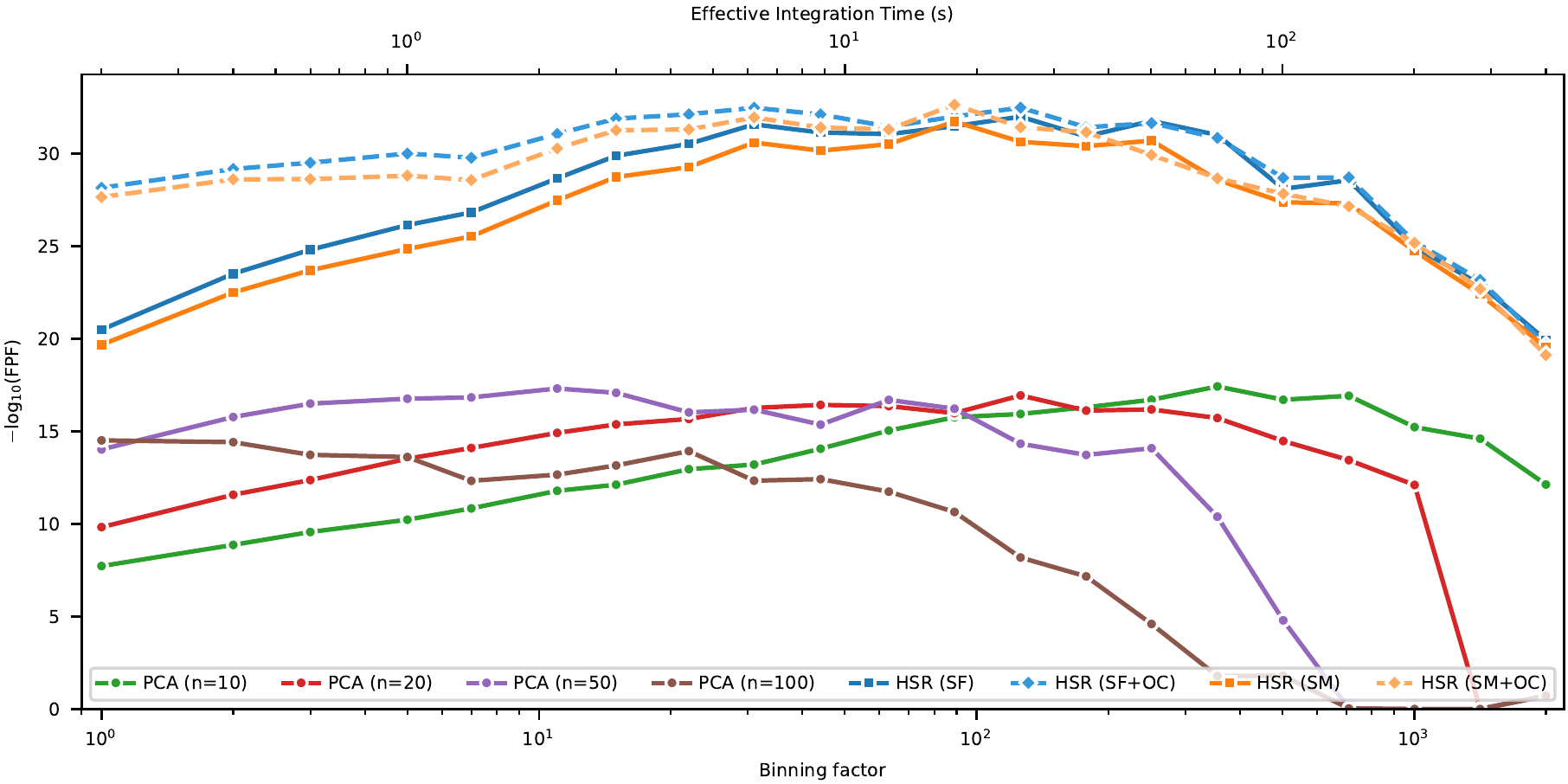}
    \caption{
        Results of experiment \ref{subsec:fpf-as-function-of-temporal-binning}:
        We plot the \logfpf score as a function of the temporal binning factor (or, equivalently, the effective integration time) for both PCA-based PSF subtraction and the two versions of our approach, signal fitting (SF) and signal masking (SM).
        For the latter, we also compare the performance with and without the observing conditions (OC) as additional predictors.
        The figure is based on experiments using the Beta Pictoris~$L'$ data set.~%
        \href{https://github.com/timothygebhard/hsr4hci/blob/master/experiments/appendix/D.1_fpf-as-function-of-temporal-binning/02_plot_logfpf_over_binning_factor.py}{\LinkToCode}
    }
    \label{fig:log-fpf-over-binning-factor-beta-pic-lp}
\end{figure*}

\paragraph{Setup:}
We begin with repeating the experiments from \cref{subsec:first-results} and \cref{subsec:oc-experiment} for different temporal binning factors, which we have chosen to give uniform coverage of the binning factor in logarithmic space.
Our experiments use the Beta~Pictoris~$L'$ data set, and we compute the \logfpf score for the one known planet in there.
Besides signal fitting and signal masking with and without the observing conditions as additional predictors, we also run PCA-based PSF subtraction for different numbers of principal components: $n_\text{PC} \in \lbrace 10, 20, 50, 100 \rbrace$.
Finally, we plot the \logfpf as a function of the binning factor, or, equivalently, the effective integration time.
The results are shown in \cref{fig:log-fpf-over-binning-factor-beta-pic-lp}.

\paragraph{Results:}
For both PCA and HSR, we notice the same basic pattern: 
Initially, the \logfpf increases as a function of the temporal binning factor, suggesting that some amount of temporal binning may be helpful for post-processing ADI data.
As we continue to increase the binning factor further, the \logfpf for both types of algorithms eventually reaches a peak, after which the performance decreases again.
In the case of PCA, the position of this peak depends on the number of principal components.
For all binning factor values, the post-processing based on half-sibling regression achieves a higher \logfpf value, with the signal fitting-variant always giving slightly higher values.

Furthermore, we notice that for small binning factors, the version of HSR that uses the observing conditions as additional predictors has a clear advantage over the version that does not include these metadata.
However, as we increase the binning factor, this advantage decreases continually. 
For a binning factor greater than approximately 100, we find no clear difference between using or not using the observing conditions.
We will revisit this effect and its potential explanation in the next section.

Regarding the effect that moderately binning the data appears to improve the performance while large binning factors decrease it, we believe there are two competing mechanisms at play here.
For small binning factors, the increase in performance that we observe may be an effect of a form of overfitting: 
Effectively, we reduce the amount of data while keeping the model capacity the same, which could mean that it becomes easier for the model to reproduce the data.
For example, for PCA, a simple experiment shows that if we keep the number of principal components fixed, the fraction of the explained variance increases as a function of the binning factor.
Additionally, moderately binning the data might already remove some systematic noise that the post-processing algorithms are otherwise not able to model well.
However, temporally binning the data also discards information. 
It appears plausible that for sufficiently large binning factors, this loss of information outweighs the effect of the relative increase in model capacity, thus leading to the observed decrease in the \logfpf score.
This observation also seems to match a finding by \citet{Samland_2021} who report a decrease in performance when comparing their method on data with an integration time of \SI{16}{\second} and \SI{64}{\second} (albeit using data from a different instrument in a different wavelength regime).

\subsection{Temporal binning and observing conditions}
\label{subsec:temporal-binning-and-oc}

In \cref{subsec:oc-experiment}, we have observed that using the observing conditions as additional predictors can significantly reduce the residual noise at small separations.
However, this reduction did not seem to translate directly into improved detection limits: 
The contrast curves we computed with observing conditions did not look significantly different from those obtained without the additional predictors.
In this supplementary experiment, we study if the discrepancy can be explained as an effect of temporal binning: 
The experiments for \cref{fig:first-results} and \cref{fig:obscon-results} have used unbinned data, whereas the experiments for the computation of the detection limits used data with a binning factor of 128.

\paragraph{Setup:}
We run the vanilla version of the half-sibling regression in a small region of interest around the star, both with and without the observing conditions, and with binning factors 1 (= no binning), 10, 100, and 1000.
The general experiment parameters (\eg shape and size of predictor region, regularization, etc.) are kept the same as in \cref{subsec:first-results} and \cref{subsec:oc-experiment}, respectively.
We limit ourselves to the Beta Pictoris $M'$, since this is the data set where we visually found the biggest difference between the signal estimates with and without the observing conditions.

\begin{figure*}[p]
    \centering
    \setlength{\tabcolsep}{0mm}
    \newcommand{\rot}[1]{\adjustbox{margin=1mm}{\rotatebox[origin=c]{90}{\textbf{\small#1}}}}
    \newcommand{\bth}[1]{\adjustbox{margin=1mm}{\textbf{\small#1}}}
    \newcommand{\imginclude}[1]{\adjincludegraphics[valign=M, width=4.3cm, margin=0.9mm]{#1}}
    
    \begin{tabular}{ccccc}
        & \bth{No binning} & \bth{Binning factor 10} & \bth{Binning factor 100} & \bth{Binning factor 1000} \\
        \rot{Without OC} & 
        \imginclude{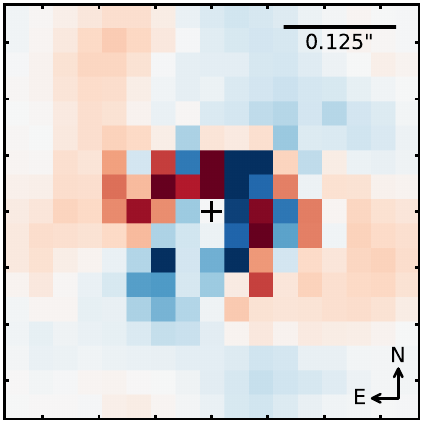} &%
        \imginclude{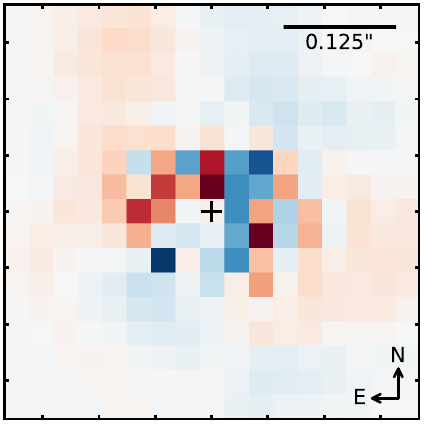} &%
        \imginclude{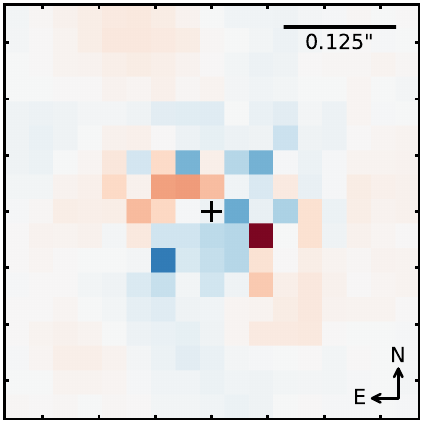} &%
        \imginclude{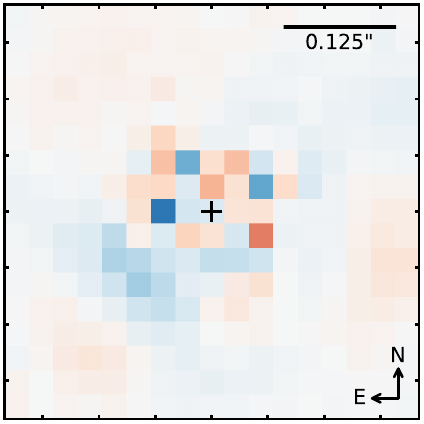} \\%
        \rot{With OC} & 
        \imginclude{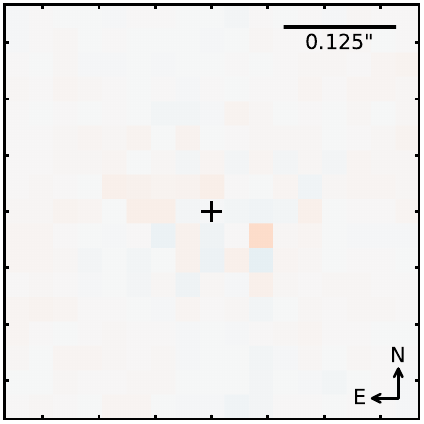} &%
        \imginclude{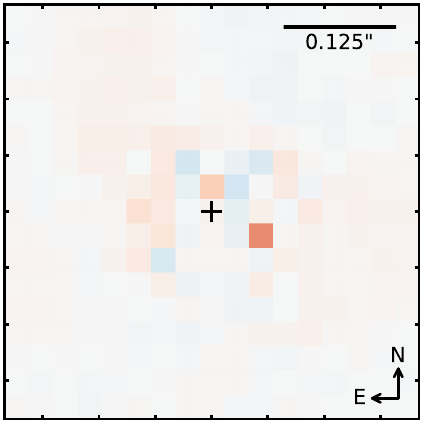} &%
        \imginclude{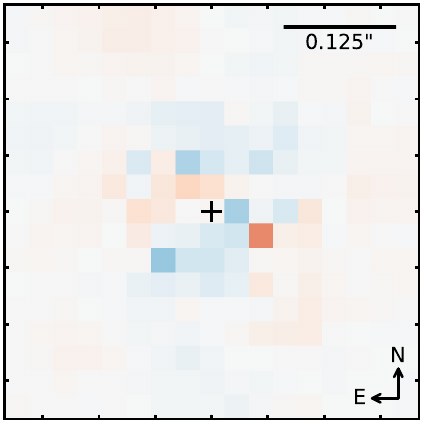} &%
        \imginclude{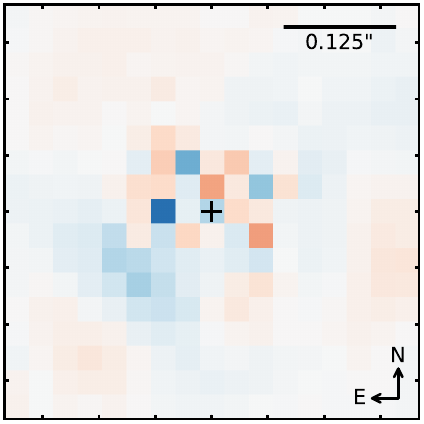} \\%
    \end{tabular}
    \caption{
        Results of experiment \ref{subsec:temporal-binning-and-oc}:        
        The figure shows the signal estimate (or rather: the residual noise present in the signal estimate) for different amounts of temporal binning and compares the results with and without the observing conditions as additional predictors.
        The results shown here are based on experiments with the Beta Pictoris $M'$ data set.
        All plots use the same scale and color bar.~%
        \href{https://github.com/timothygebhard/hsr4hci/blob/master/experiments/appendix/D.2_residual-noise-with-and-without-oc/make_plots.py}{\LinkToCode}
    }
    \label{fig:comparison-with-without-oc}
\end{figure*}

\paragraph{Results:}
Looking at the results in \cref{fig:comparison-with-without-oc}, we notice several things.
First, without observing conditions, the amount of residual noise decreases with the binning factor.
Second, as we increase the binning factor, the difference between the results with and without observing conditions becomes smaller and smaller:
Without temporal binning, adding the observing conditions as predictors results in significantly less residual noise at small separations, whereas for binning factors beyond 100, the signal estimates with and without observing conditions become virtually indistinguishable.
Both of these effects seem consistent with our observations in \cref{subsec:fpf-as-function-of-temporal-binning}, and like in our explanation there, we conjecture that this effect can be explained by the fact that temporal binning essentially discards information and thus reduces the usefulness of the additional predictors.

We conclude from this experiment and the previous one that the observing conditions can improve the denoising of HCI data; however, the temporal resolution appears to be of critical importance here.
Furthermore, we find that the results of this experiment provide a sufficient explanation of the effect mentioned in the introduction of this section:
Since our contrast curves in \cref{subsec:oc-experiment} were computed from experiments using data with a binning factor of 128, we are not surprised that in this case, adding the observing conditions does not improve the detection limits. 
 
\end{document}